\pdfoutput=1
\documentclass[usenatbib]{mnras}
\usepackage{graphicx}
\usepackage{color}
\usepackage{hyperref,breakurl}
\usepackage[usenames,dvipsnames]{xcolor}
\usepackage{dcolumn}
\usepackage{bm}
\usepackage{hyperref}
\usepackage{array}
\usepackage{dcolumn}
\usepackage{tabularx}
\usepackage{array}
\newcolumntype{L}{>{\centering\arraybackslash}m{0.45\columnwidth}}

\usepackage{amsmath,amssymb,latexsym,times}

\usepackage[normalem]{ulem}
\defcitealias{2009arXiv0912.0201L}{LSST Science Book} 
\defcitealias{LPM-17}{LSST SRD} 
\defcitealias{2018arXiv180901669T}{DESC SRD} 
\defcitealias{2019ApJS..245...26K}{CosmoDC2} 
\defcitealias{2021ApJS..253...31L}{DC2ImageSim} 
\defcitealias{2021arXiv210104855L}{DC2DRNote} 

\usepackage{lineno}

\title[A Joint Roman--Rubin Synthetic Wide-Field Imaging Survey]{A Joint Roman Space Telescope and Rubin Observatory Synthetic Wide-Field Imaging Survey}

\author[Troxel et al.]{
\parbox{\textwidth}{
\Large
M.~A.~Troxel,$^{1}$
C.~Lin,$^{1}$
A.~Park,$^{2}$
C.~Hirata,$^{3,4}$
R.~Mandelbaum,$^{2}$
M.~Jarvis,$^{5}$
A.~Choi,$^{6}$
J.~Givans,$^{7}$
M.~Higgins,$^{1}$
B.~Sanchez,$^{1}$
M.~Yamamoto,$^{1}$
H.~Awan,$^{8}$
J.~Chiang,$^{9}$
O.~Dor\'e,$^{10,11}$
C.~W.~Walter,$^{1}$
T.~Zhang,$^{2}$
J.~Cohen-Tanugi,$^{12,13}$ 
E.~Gawiser,$^{14}$
A.~Hearin,$^{15}$
K.~Heitmann,$^{15}$
M.~Ishak,$^{16}$
E.~Kovacs,$^{15}$
Y.-Y.~Mao,$^{17}$
M.~Wood-Vasey$^{18}$
and the LSST Dark Energy Science Collaboration
}
\vspace{0.4cm}
\\
\parbox{\textwidth}{
$^{1}$ Department of Physics, Duke University, Durham, NC 27708, USA\\
$^{2}$ Carnegie Mellon University, 5000 Forbes Ave, Pittsburgh, PA 15213, USA\\
$^{3}$ Department of Physics, The Ohio State University, 191 West Woodruff Ave, Columbus, Ohio 43210, USA\\
$^{4}$ Department of Astronomy, The Ohio State University, 140 West 18th Ave, Columbus, Ohio 43210, USA\\
$^{5}$ University of Pennsylvania, Philadelphia, PA 19104, USA\\
$^{6}$ California Institute of Technology, 1200 E California Blvd, Pasadena, CA 91125, USA\\
$^{7}$ Department of Astrophysical Sciences, Princeton University, 4 Ivy Lane, Princeton, NJ 08540, USA\\
$^{8}$ Leinweber Center for Theoretical Physics, Department of Physics, University of Michigan, Ann Arbor, MI 48109, USA\\
$^{9}$ SLAC National Accelerator Laboratory, 2575 Sand Hill Road, Menlo Park, CA, 94025, USA\\
$^{10}$ California Institute of Technology, 1200 E. California Boulevard, Pasadena, CA 91125, USA\\
$^{11}$ Jet Propulsion Laboratory, California Institute of Technology, 4800 Oak Grove Drive, Pasadena, CA 91109, USA\\
$^{12}$ Laboratoire Univers et Particules de Montpellier, Place Eug\`ene Bataillon - CC 72, F-34095 Montpellier Cedex 05, France\\
$^{13}$ LPC, université Clermont Auvergne, CNRS, F-63000 Clermont-Ferrand, France\\
$^{14}$ Department of Physics and Astronomy, Rutgers University, Piscataway, NJ 08854, USA\\
$^{15}$ Argonne National Laboratory, 9700 S Cass Ave, Lemont, IL 60439, USA\\
$^{16}$ University of Texas, Dallas, 800 W Campbell Rd, Richardson, TX 75080, USA\\
$^{17}$ University of Utah, Department of Physics and Astronomy 115 South 1400 East, Salt Lake City, UT 84112-0830, USA\\
$^{18}$ University of Pittsburgh, 4200 Fifth Ave, Pittsburgh, PA 15260, USA\\
\vspace{-0.7cm}
}
}

\pubyear{2022}

\begin{document}
\label{firstpage}
\pagerange{\pageref{firstpage}--\pageref{lastpage}}
\maketitle

\begin{abstract}
We present and validate 20 deg$^2$ of overlapping synthetic imaging surveys representing the full depth of the Nancy Grace Roman Space Telescope High-Latitude Imaging Survey (HLIS) and five years of observations of the Vera C. Rubin Observatory Legacy Survey of Space and Time (LSST).
The two synthetic surveys are summarized, with reference to the existing 300 deg$^2$ of LSST simulated imaging produced as part of Dark Energy Science Collaboration (DESC) Data Challenge 2 (DC2). Both synthetic surveys observe the same simulated DESC DC2 universe. For the synthetic Roman survey, we simulate for the first time fully chromatic images along with the detailed physics of the Sensor Chip Assemblies derived from lab measurements using the flight detectors. The simulated imaging and resulting pixel-level measurements of photometric properties of objects span a wavelength range of $\sim$0.3 to 2.0~$\mu$m. We also describe updates to the Roman simulation pipeline, changes in how astrophysical objects are simulated relative to the original DC2 simulations, and the resulting simulated Roman data products. We use these simulations to explore the relative fraction of unrecognized blends in LSST images, finding that 20-30\% of objects identified in LSST images with $i$-band magnitudes brighter than 25 can be identified as multiple objects in Roman images. These simulations provide a unique testing ground for the development and validation of joint pixel-level analysis techniques of ground- and space-based imaging data sets in the second half of the 2020s -- in particular the case of joint Roman--LSST analyses.
\end{abstract}


\maketitle

\section{Introduction}\label{sec:intro}

The 2020s will be a remarkable decade for observational cosmology and astrophysics as we gain access to data from several new observatories and experiments that will largely supersede in their first years what is possible with current experiments like the Dark Energy Survey (DES),\footnote{\url{https://www.darkenergysurvey.org}} the Hyper-Suprime Cam Strategic Survey Program (HSC),\footnote{\url{https://hsc.mtk.nao.ac.jp/ssp/}} the Kilo-Degree Survey (KiDS),\footnote{\url{https://kids.strw.leidenuniv.nl}} and the Sloan Digital Sky Survey (SDSS).\footnote{\url{https://www.sdss.org}} These new observatories cover both photometric and spectroscopic observations with a significantly faster cadence and greater sensitivity of observation than is possible with current experiments. They include the ground-based Dark Energy Spectroscopic Instrument (DESI)\footnote{\url{https://www.desi.lbl.gov}} and Vera C.\ Rubin Observatory\footnote{\url{https://www.lsst.org}} (photometric) and the space-based Euclid\footnote{\url{https://sci.esa.int/web/euclid}} and Nancy Grace Roman Space Telescope,\footnote{\url{https://roman.gsfc.nasa.gov}} which have both  photometric and spectroscopic capabilities. The combination of deep photometric and spectroscopic observations from both ground- and space-based telescopes will provide new capabilities to overcome limitations of any single experiment by combining  observations from multiple observatories \citep{2017ApJS..233...21R,2019arXiv190410439C}.

The survey missions of these observatories all include a focus on wide-field static cosmology research -- in particular, the study of dark matter and dark energy. In this paper, we focus on simulating the imaging components of two of these surveys. For the Roman Space Telescope, this is the High-Latitude Imaging Survey \citep[HLIS;][]{2015arXiv150303757S,2019arXiv190205569A}, and for the Rubin Observatory, this is the Legacy Survey of Space and Time \citep[LSST;][]{2019ApJ...873..111I}.  While both surveys will provide ground-breaking results on their own, each has limitations that can be further overcome when analyzed jointly  \citep{2019BAAS...51g.202C,2020arXiv200810663C,Eifler2019Partnering,2019BAAS...51c.201R,2021MNRAS.507.1514E}. From the ground, LSST is faced with the challenge of deblending the light profiles of objects convolved with the atmospheric point-spread function (PSF). The higher resolution of the Roman images, both in pixel scale and PSF, can be used to improve the deblending of these ground-based images \citep{2019arXiv190108586S,2021arXiv210706984J,Melchior_2021}. Both surveys also benefit from a wider optical-to-near-infrared  filter passband coverage than either individually (approximately 0.3 to 2.0 $\mu$m) to improve the reliability of photometric redshifts (photo-$z$s) \citep{graham2020}. Roman also offers a spectroscopic emission-line object survey for further redshift calibration. Shallower spectroscopic and near-infrared data overlapping with the LSST can also be obtained in the first years of LSST from DESI \citep{2016arXiv161100036D} and Euclid \citep{2011arXiv1110.3193L}, respectively. All-sky near-IR spectroscopy will also be available from SphereX \citep{2014arXiv1412.4872D}.\footnote{\url{https://spherex.caltech.edu}}

The Roman HLIS Reference Survey (defined further in Sec.~\ref{surveys-roman}) will nominally observe for five years (approximately 2027--2032) in the Y106/J129/H158/F184 bandpasses\footnote{Lowercase letters ($ugrizy$) will be used to refer to Rubin bandpasses throughout, while uppercase will refer to Roman (Y106/J129/H158/F184).} in the near-infrared over a 2000 deg$^2$  region of the sky that is fully contained within the LSST footprint. The HLIS is designed to target weak lensing, galaxy clustering, and galaxy cluster counts to probe the evolution of large-scale structure in the Universe. It will observe to a 5$\sigma$ point-source limiting magnitude of almost 27 in Y106/J129/H158 bands \citep{2015arXiv150303757S}. Images will be undersampled, with a pixel scale of 0.11 arcsec and PSF FWHM of 0.27 arcsec in the H158 band. The Roman Wide-Field Instrument has 18 4096$\times$4096 Sensor Chip Assemblies (SCAs), with a total field-of-view of 0.281 deg$^2$. More details about the Roman Wide Field Instrument are available on the Roman Project website.\footnote{\url{https://roman.gsfc.nasa.gov/science/technical_resources.html}} The Reference survey will observe about 6 overlapping translationally and rotationally dithered images using each of the four filters. The final HLIS observing strategy has not yet been selected, and will be decided based on a community-informed process over the next few years. For this paper, we simulate the HLIS Reference Survey.

LSST is a photometric survey, designed to observe about 20,000 deg$^2$ of the southern sky for 10 years (approximately 2024--2034) in $ugrizy$ bands (\citet{2009arXiv0912.0201L}; hereafter referred to as the LSST Science Book). The survey is divided into two components: 1) a main, wide-field survey (wide-fast-deep, or WFD) comprising at least 18,000 deg$^2$ over 10 years, with a median of 825 visits to each field and median single-visit depths for $ugrizy$ = (23.9, 25.0, 24.7, 24.0, 23.3, 21.7) (LSST Science Requirements Document, \citet{LPM-17}; hereafter referred to as the LSST SRD), and 2) mini-surveys, including Deep Drilling Fields (DDFs), four of which are formally announced by the Rubin Observatory Project.\footnote{\url{https://www.lsst.org/scientists/survey-design/ddf}} While LSST goals range from studying near-Earth objects to transient phenomena \citepalias{2009arXiv0912.0201L}, the WFD survey is particularly suited for dark energy probes, including galaxy clustering, weak lensing, clusters, Type Ia supernovae, and strong lensing (LSST Dark Energy Science Collaboration (DESC)\footnote{\url{http://lsstdesc.org}}  Science Requirements Document, \citet{2018arXiv180901669T}; hereafter referred to as the DESC~SRD); the DDFs, on the other hand, will provide deeper, more complete samples for multiple purposes such as photo-$z$ calibration and transients templates. While the exact observing strategy and footprint are currently being  optimized for scientific gain,\footnote{See more at \url{https://www.lsst.org/content/charge-survey-cadence-optimization-committee-scoc} and \citet{2022ApJS..258....1B}.} we have access to detailed simulations of the full baseline survey, accounting for realistic factors including and beyond the scheduling of observations, telescope pointing, slewing and downtime, and site conditions \citep{delgado+2014,2016SPIE.9910E..13D,2019AJ....157..151N}. In this work, we focus specifically on the WFD survey; we describe the exact observing cadence used for the DC2 simulation in Sec.~\ref{sec:rubinwfd}.

To take advantage of these future gains from joint-pixel analysis of Roman HLIS and LSST data, we must have realistic synthetic survey data on which to develop and test algorithms and analysis strategies \citep{2019BAAS...51g.210P}. This paper builds on simulation efforts from the LSST DESC, including \cite{2019ApJS..245...26K} (hereafter CosmoDC2), \cite{2021ApJS..253...31L} (hereafter DC2ImageSim), and \cite{2021arXiv210104855L} (hereafter DC2DRNote) that describe the existing LSST synthetic survey used in this paper and the Roman Cosmology with the HLS Science Investigation Team (SIT) \citep{2021MNRAS.501.2044T} to produce the first large-scale realistic joint-survey image simulation. The simulated imaging data provide an overlapping simulated survey region of both the LSST WFD survey and Roman HLIS surveys covering approx.~20 deg$^2$ of the same, realistic simulated sky. The simulated image products include the first five years of  LSST and the full five years of Roman HLIS data. The new Roman component of the simulation was produced for this work and currently focuses on static phenomena, but future versions will contain transient phenomena that the two surveys can jointly study during the full five year temporal overlap during the Roman HLIS primary mission. While targeted to the Roman and Rubin observatories, these simulated data products provide a realistic testing ground for development of algorithms for combining any deep ground- and space-based observatories (e.g., Euclid; \citealt{2011arXiv1110.3193L}).

These simulations enable pixel-level measurements of photometric properties of the simulated objects spanning $\sim$0.3 to 2.0~$\mu$m. We validate these photometric properties against existing predictions from Hubble Space Telescope observations for the Rubin LSST and Roman HLIS. The joint synthetic surveys are used to explore the relative fraction of unrecognized blends in LSST images, finding that 20-30\% of objects identified in LSST images with $i$-band magnitudes brighter than 25 can be identified as multiple objects in Roman images. The joint synthetic surveys presented provide a unique testing ground for the development and validation of joint pixel-level analysis techniques of ground- and space-based imaging data sets -- in particular the case of joint Roman--LSST analyses in the second half of the 2020s.

We describe in Sec.~\ref{dc2} the simulated universe of the LSST DESC DC2. In Sec.~\ref{surveys}, the wide-field Roman HLIS and LSST surveys are described. The image simulation process is described in Sec.~\ref{sec:imagesims}, and in Sec.~\ref{results} we describe the simulation output, validation, and initial studies of blending in the two surveys. We conclude in Sec.~\ref{conclusions}.


\begin{figure}
\begin{center}
\begin{tabular}{LL}
  \includegraphics[width=0.48\columnwidth]{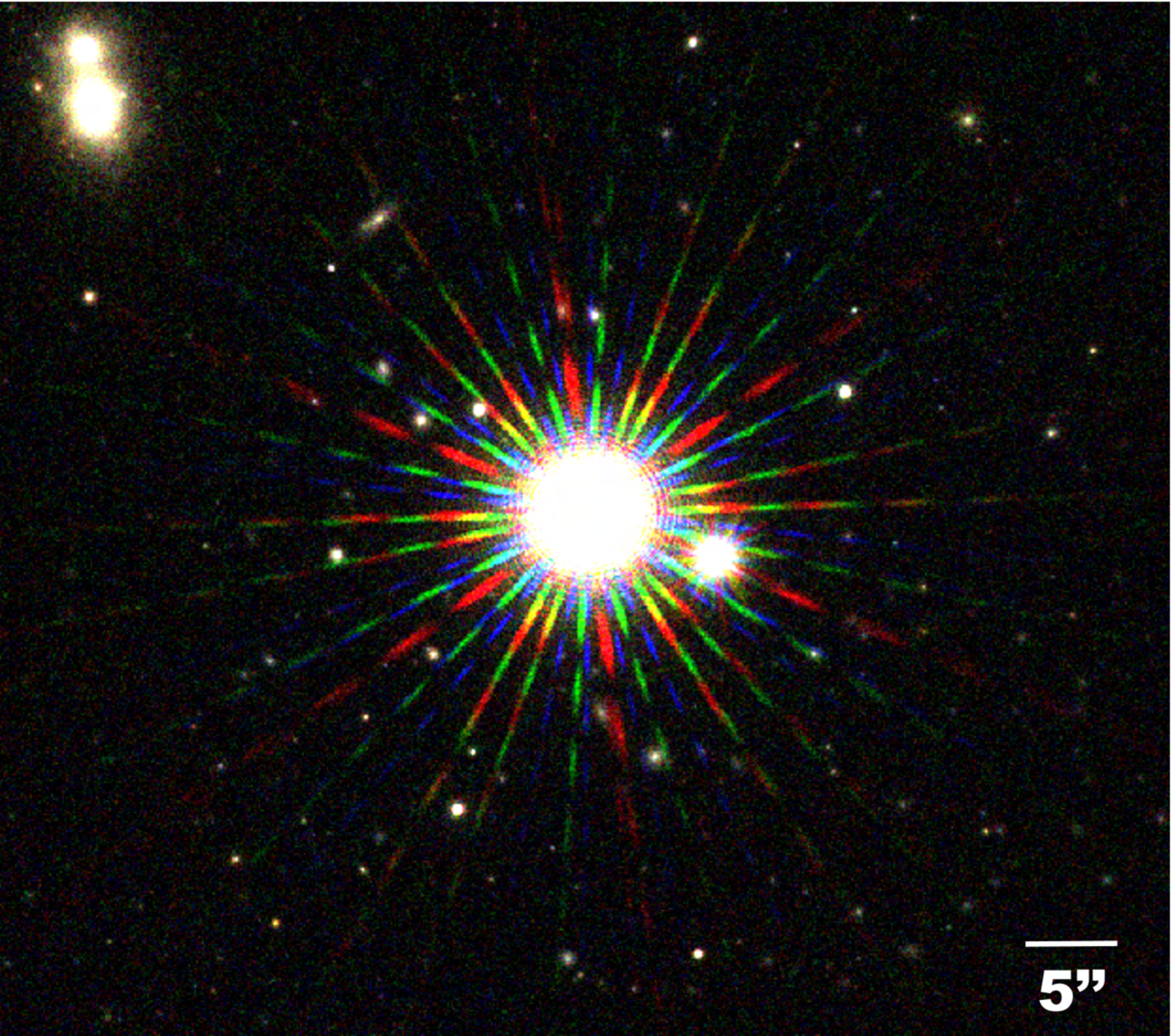} &   \includegraphics[width=0.48\columnwidth]{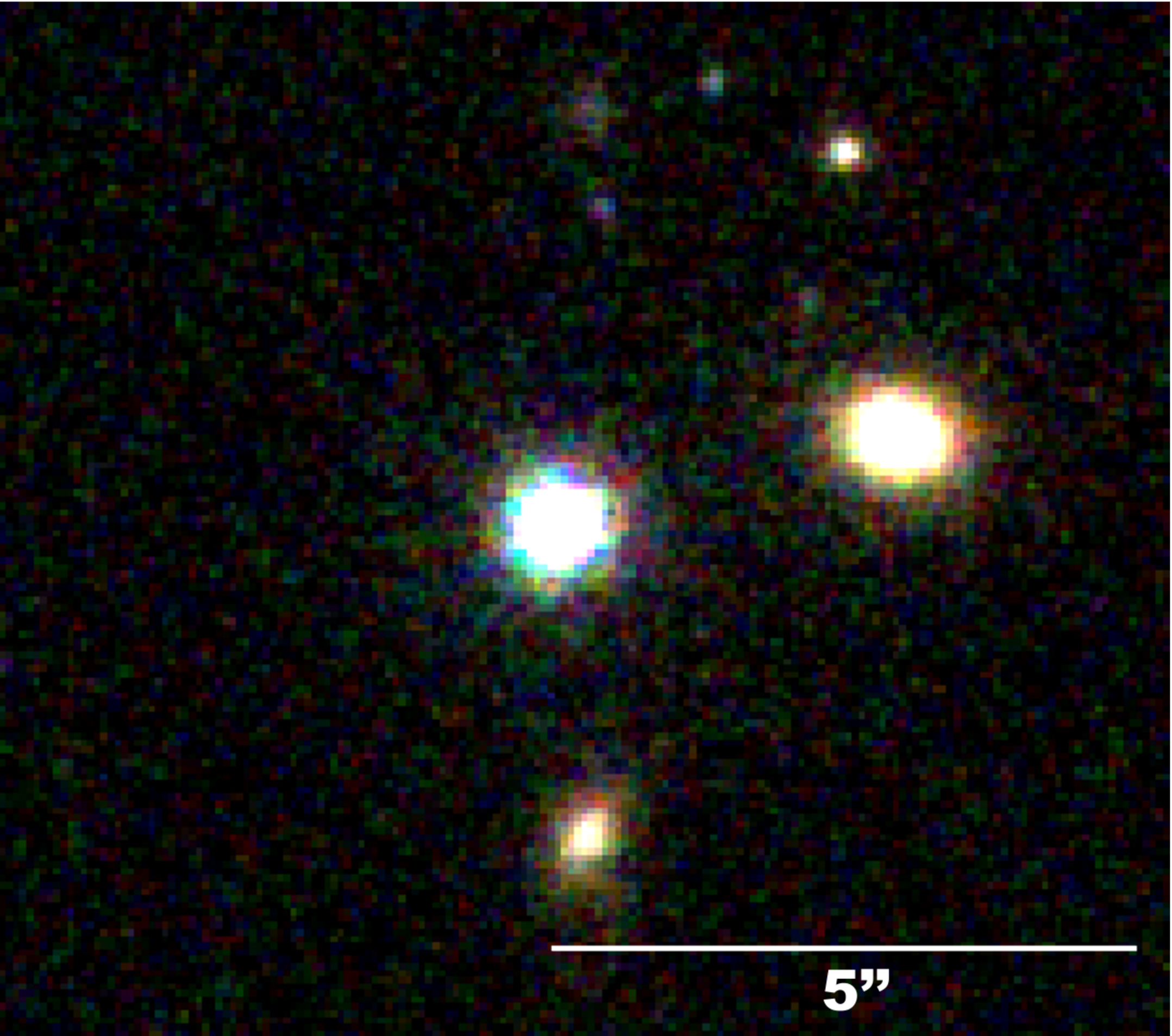} \\
(a) Bright star & (b) Faint star \\[6pt]
  \includegraphics[width=0.48\columnwidth]{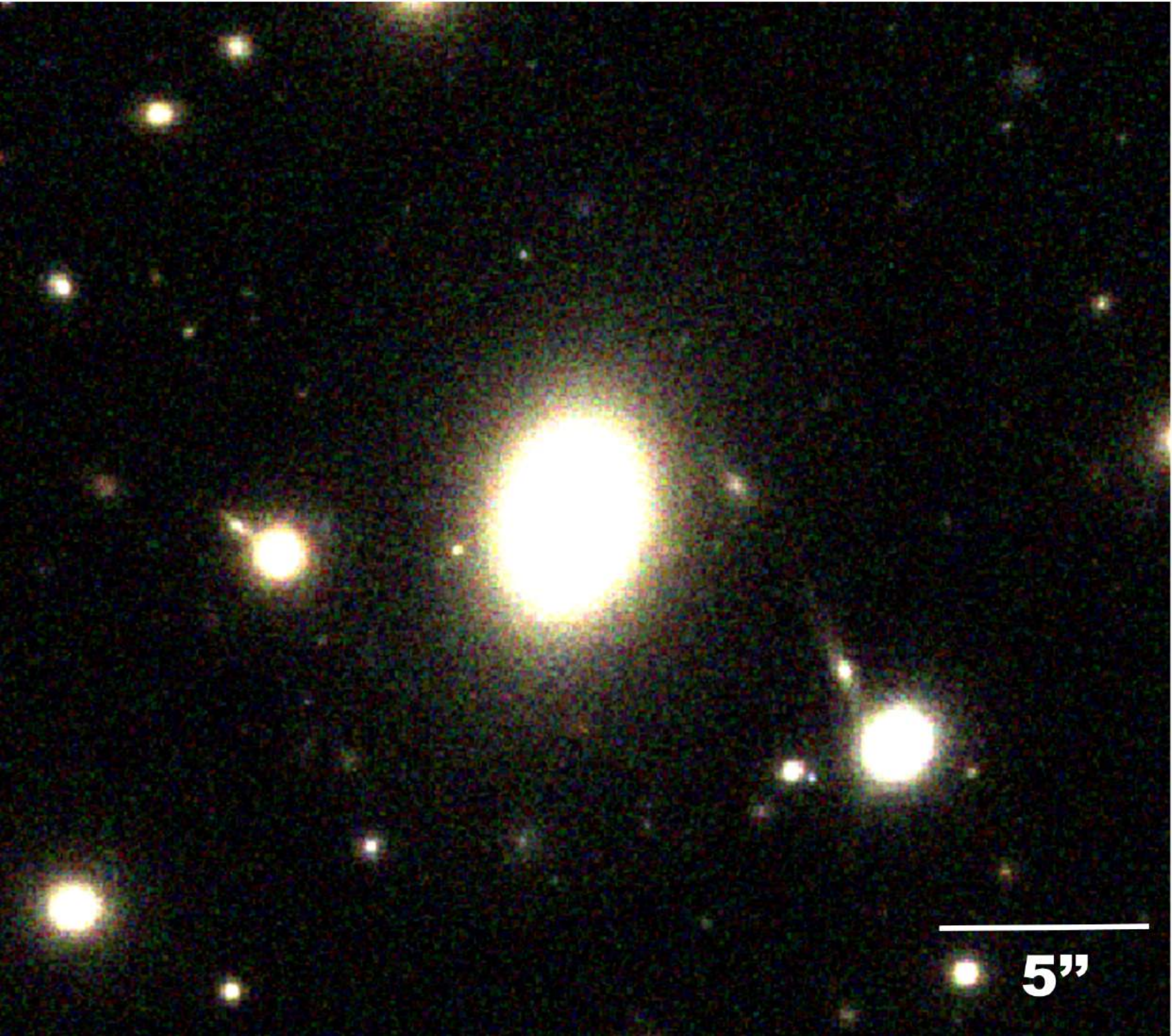} &   \includegraphics[width=0.48\columnwidth]{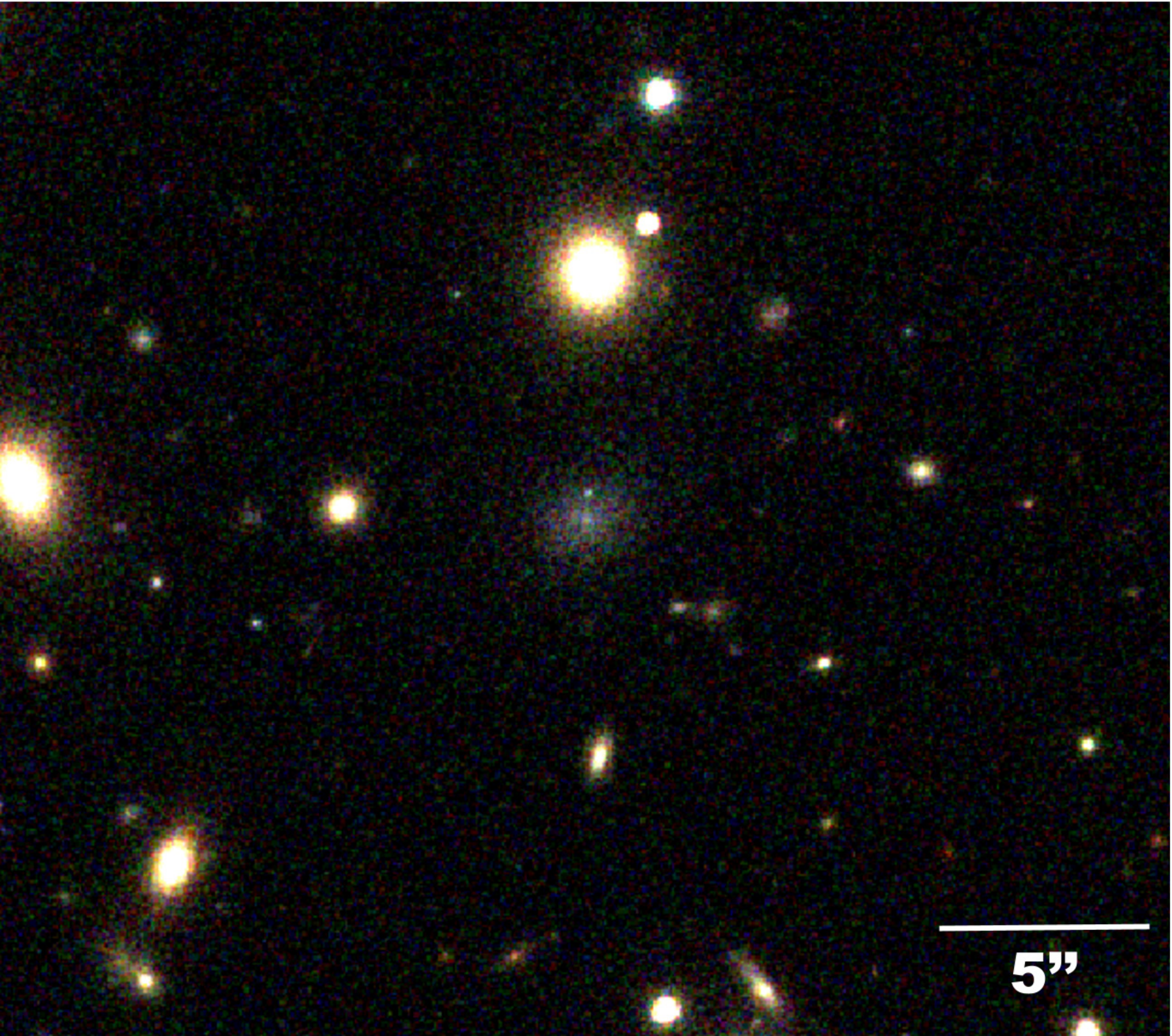} \\
(c) Bright bulge-dominated galaxy & (d) Faint bulge-dominated galaxy \\[6pt]
  \includegraphics[width=0.48\columnwidth]{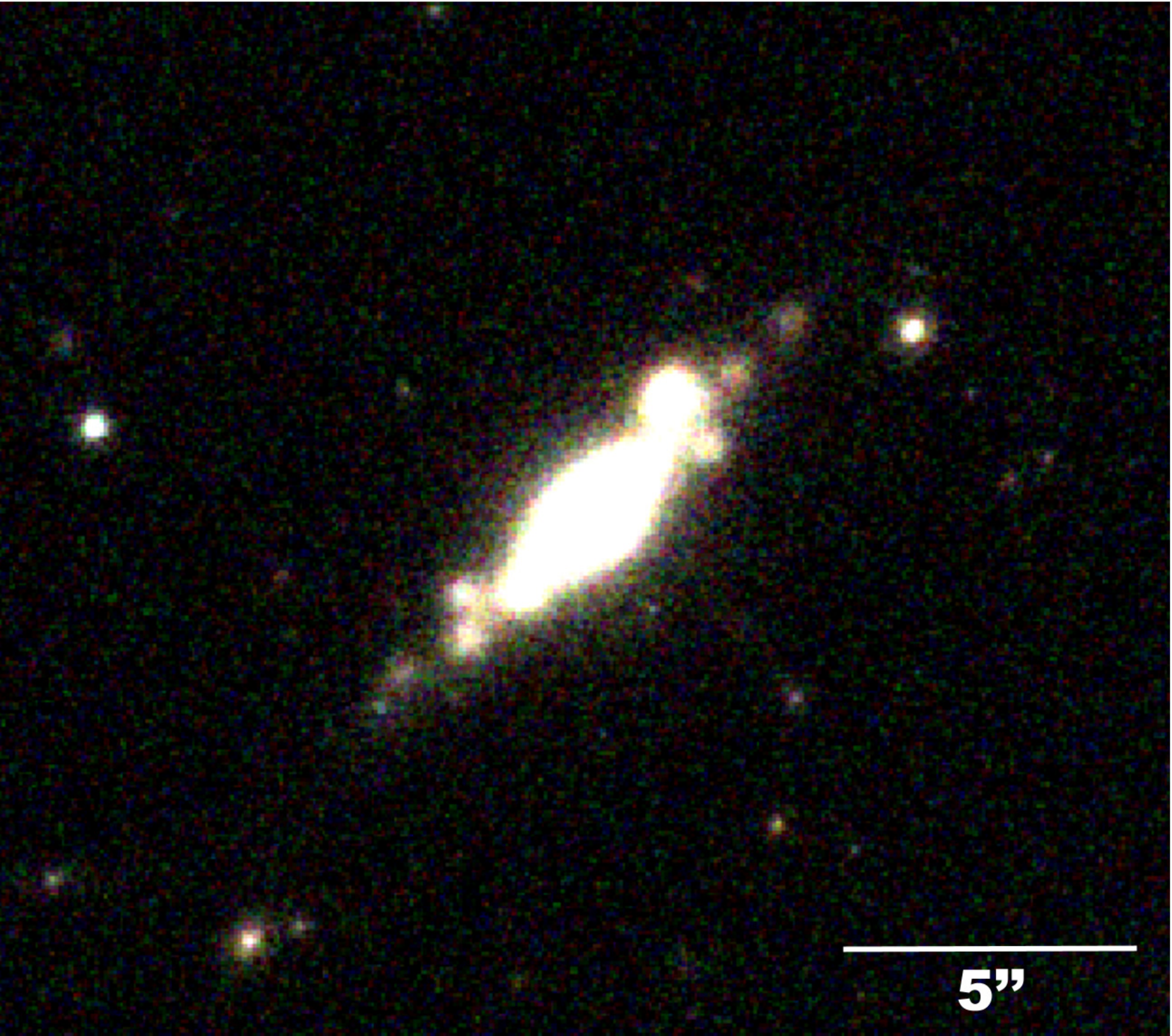} &   \includegraphics[width=0.48\columnwidth]{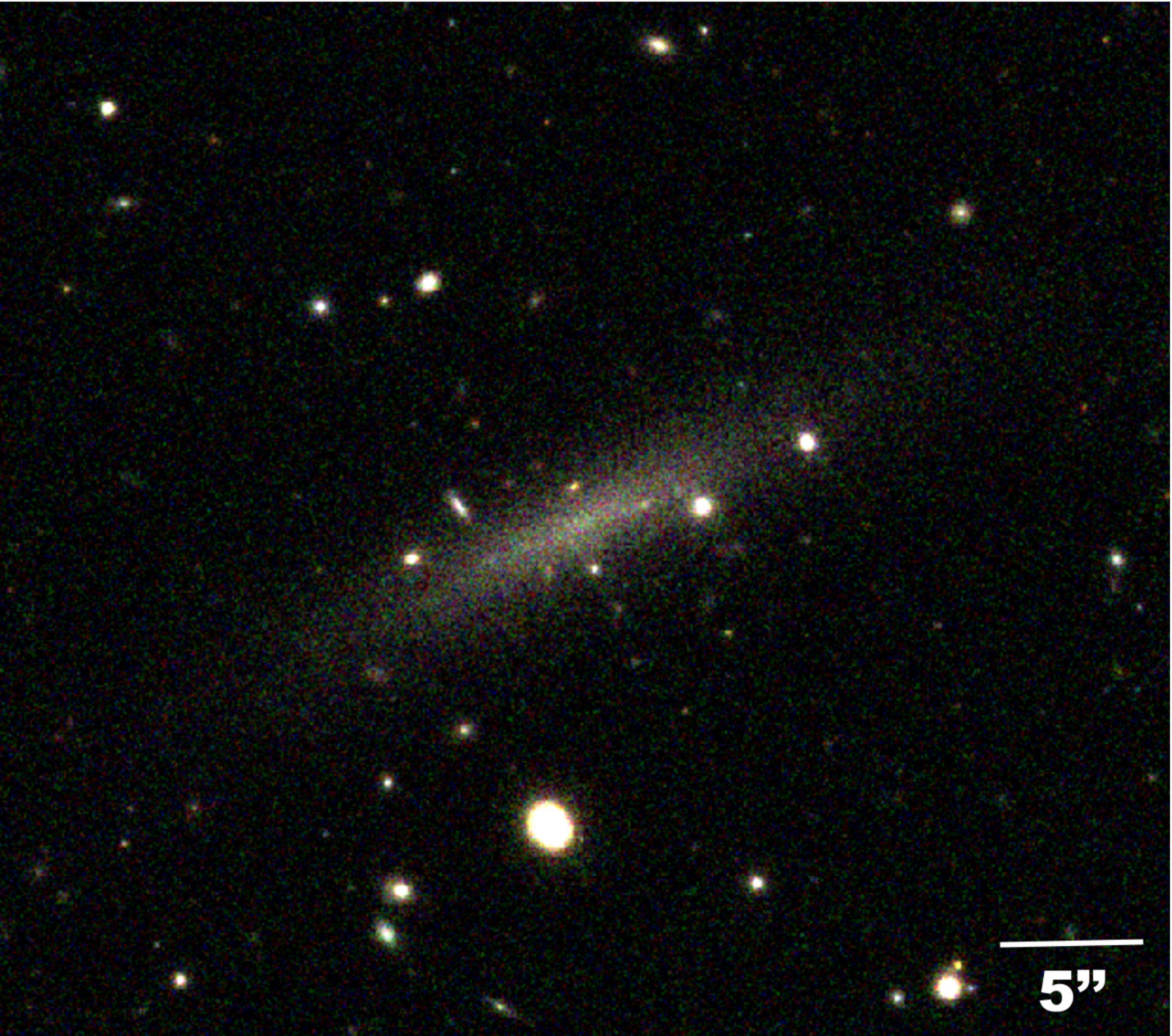} \\
(e) Bright disk-dominated galaxy with star-forming knots & (f) Faint disk-dominated galaxy \\[6pt]
  \includegraphics[width=0.48\columnwidth]{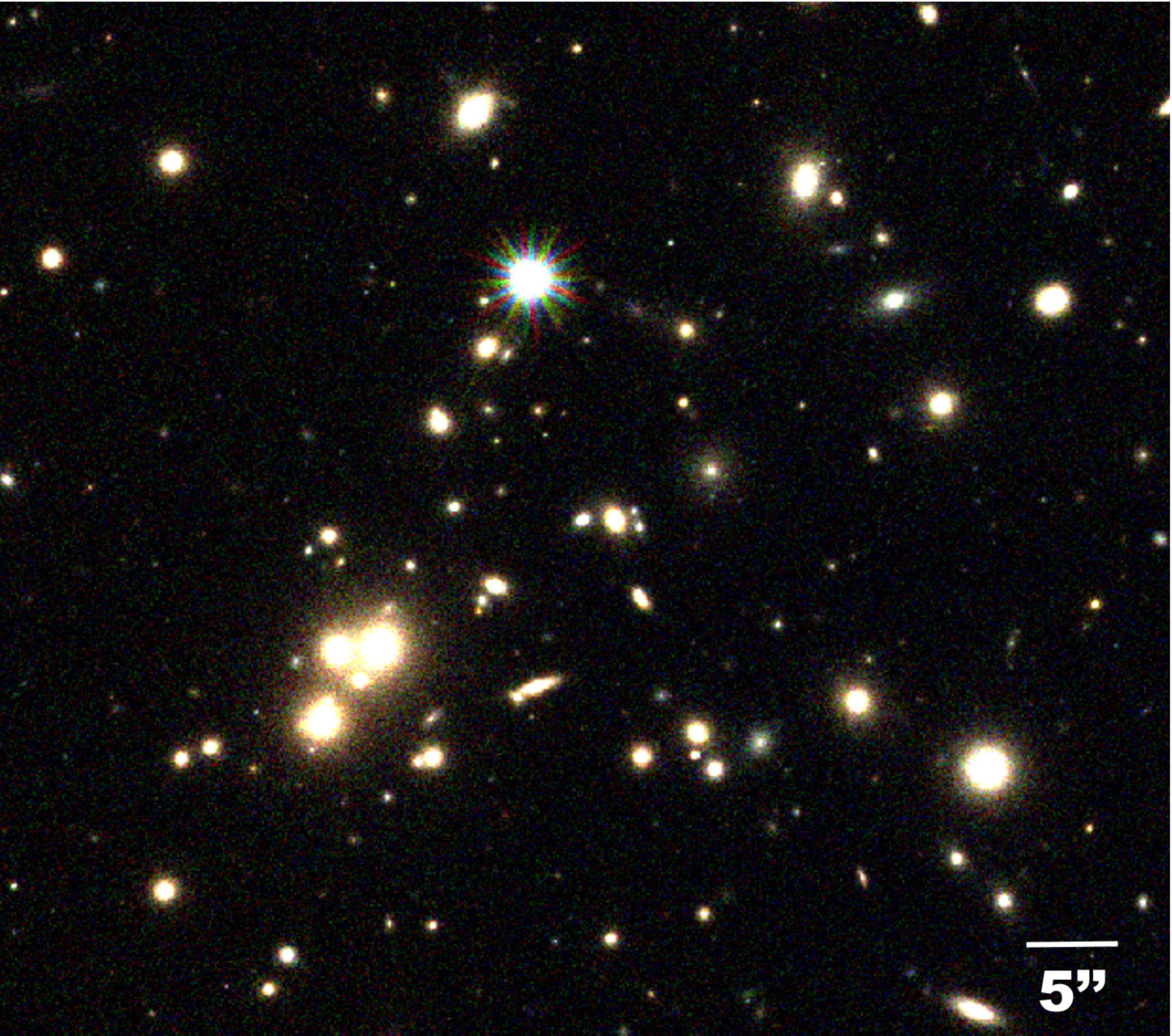} &   \includegraphics[width=0.48\columnwidth]{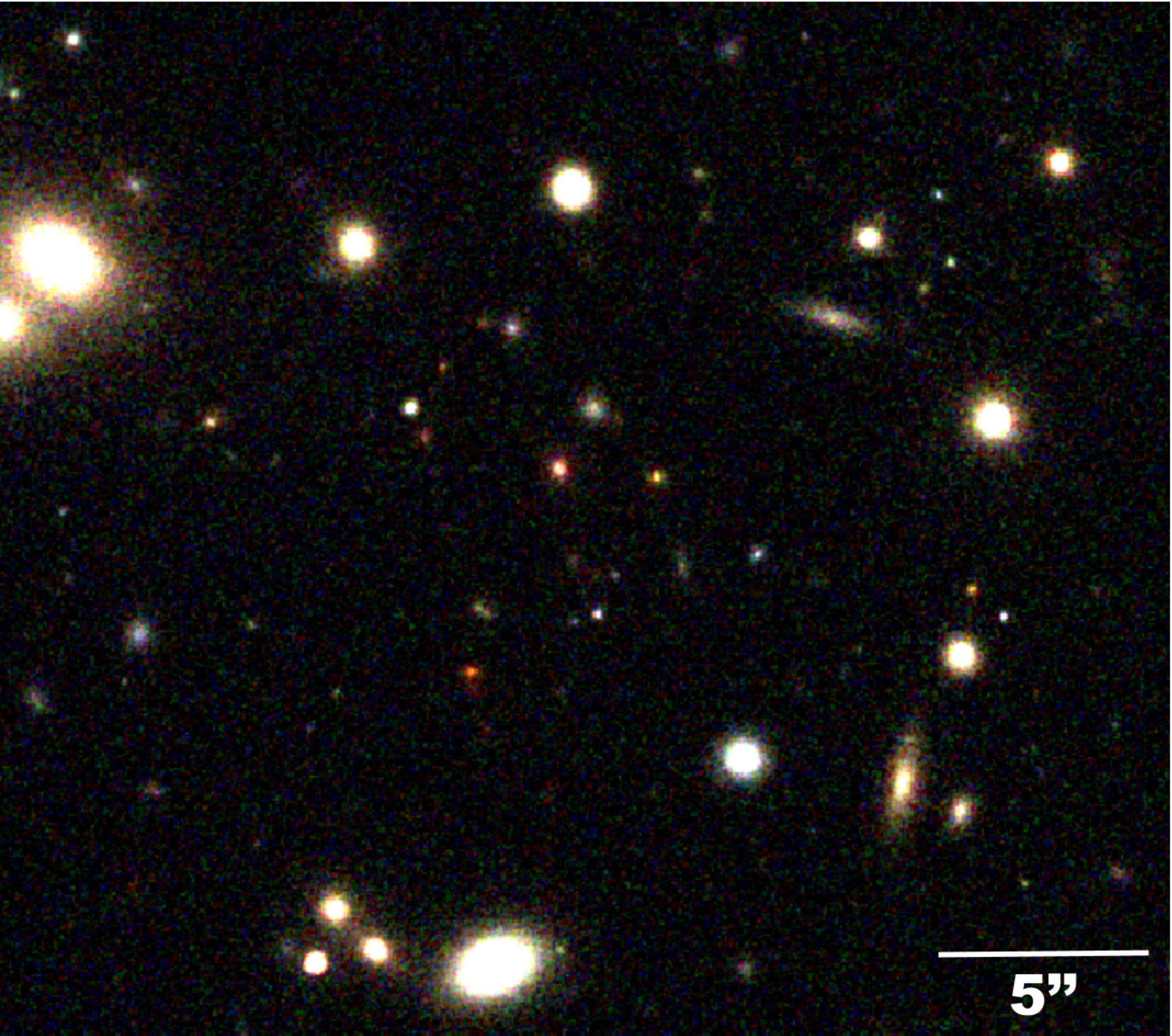} \\
(g) Bright, dense scene of galaxies & (h) Fainter scene of galaxies with several high-redshift objects and color variety \\[6pt]
\end{tabular}
\end{center}
\caption[]{Some example Y106/J129/H158-color cutouts of star, bulge- and disk- dominated galaxies, and larger scenes from the simulated five-year Roman HLIS coadd images.
\label{fig:cutouts}}
\end{figure}

\begin{figure*}
\begin{center}
\includegraphics[width=\textwidth]{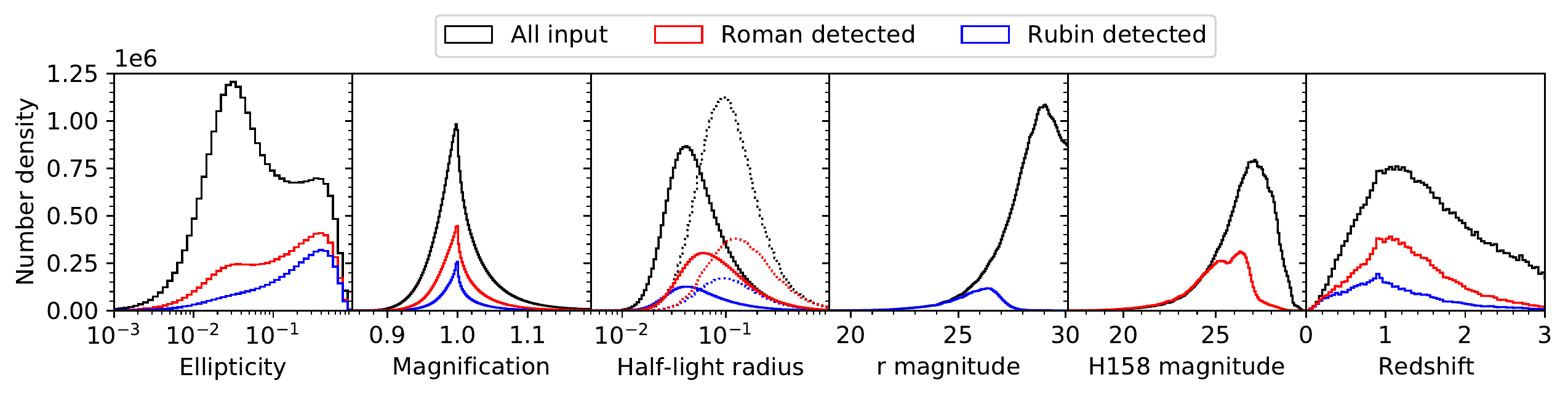}
\end{center}
\caption[]{A comparison of the input distributions of several galaxy properties are shown in black, split by component where applicable -- bulge (solid) and disk (dotted). The red histograms show the distributions of true properties of objects detected in the five-year Roman HLIS coadd images, while blue histograms show the true properties of detected LSST objects.
\label{fig:lensprop}}
\end{figure*}

\section{LSST DESC DC2 Universe}\label{dc2}

Here we briefly summarize the inputs to the LSST DC2 image simulation, and
refer readers to \citetalias{2019ApJS..245...26K}
for a more thorough description of the input extragalactic catalog (CosmoDC2),
to \citet{2022OJAp....5E...1K} for how that catalog was validated, and
to \citetalias{2021ApJS..253...31L} for other inputs.

The CosmoDC2 extragalactic catalog (\citetalias{2019ApJS..245...26K}, \citet{2022OJAp....5E...1K}) was produced
using a data-driven approach to semi-analytic modeling of the galaxy population, applied to the large-volume `Outer Rim'
simulation \citep{2019ApJS..245...16H}.  By using empirical methods for comparing the simulated
galaxy population with real observational datasets, the results of the semi-analytic model
Galacticus \citep{2012NewA...17..175B} were resampled \citep{2020MNRAS.495.5040H} to provide a galaxy population that
met the simulation specifications for use of the catalog for cosmological science cases, including
realistic effects such as variation in galaxy properties with redshift, and an appropriate level of
galaxy blending.  Many of the
quantities needed to generate image simulations were produced as a part of this process, with
galaxies represented as a sum of a galaxy bulge and disk, each with a spectral energy distribution (SED) provided using a set of
30 top-hat filters spanning ultraviolet to infrared wavelengths.

As described in Sec.~5.1.1 of \citetalias{2021ApJS..253...31L}, the input catalogs with bulge and disk
S\'ersic profile parameters \citep{sersic1963influence} specified were modified to move some of the flux from the disk into a
random walk component.  This new component adds some number of equal-flux point sources with
the same SED as the disk component, which is drawn from a Gaussian spatial distribution matching the
size and shape of the
disk.  The goal of adding this component was to roughly emulate the appearance of star-forming
regions within galaxy disks, thereby incorporating some of the realistic complexity in disk galaxy
light profiles beyond a simple S\'ersic profile.

Other extragalactic inputs to DC2 beyond those in CosmoDC2 are the time-domain elements described in
Sec.~5.2 of \citetalias{2021ApJS..253...31L}: Type Ia supernovae, strongly lensed supernovae, active
galactic nuclei (AGN), strongly lensed AGN, and strongly lensed galaxies.  However, in the area
overlapping the Roman HLIS simulations, only the Type Ia supernovae were included.  We refer readers to
that work for more details on their implementation.

Finally, the stellar population in the DC2 simulation is from Galfast \citep{2008ApJ...673..864J},
which is based on an extrapolation of the observed SDSS stellar catalogs to lower luminosities.
Further details of the stellar population modeling, including how variability was incorporated, can
be found in Sec.~5.3 of \citetalias{2021ApJS..253...31L}.

We show some examples of the diversity of the realized DC2 object profiles as simulated in the Roman HLIS  images in Fig.~\ref{fig:cutouts}. Figure \ref{fig:lensprop} shows the input distributions of several relevant object properties, such as ellipticity, magnification, half-light radius, magnitude, and redshift. There is a known issue in the DC2 object properties that leads to too many round objects \citep{eve_kovacs_2022_6336066}.

\subsection{Modifications for use in Roman HLIS images}

Our focus for the simulation on the Roman side has been static wide-field studies, so we have not implemented transient or variable sources in the Roman HLIS images. This includes supernovae and variable stars -- any variable stars have a constant magnitude selected from an arbitrary time. We have also needed to modify the input information for use in Roman HLIS images in some small but potentially impactful ways that should be considered in any quantitative analysis. First, instead of the top-hat bin reweighting of the SEDs used to produce the observed fluxes in the Rubin filter passbands, we directly use the SEDs as modeled in CosmoDC2. The top-hat bandpasses were developed to improve the modeling of optical colors for Rubin, but the Roman bandpasses extend slightly redder than the edge of the top-hat SEDs. This will introduce a modest discontinuity in the object SEDs between the optical and near-infrared, which will need to be modeled in any precise photometric template comparison between the surveys.

Second, the approach of modeling star-forming knots as a distribution of discrete point sources used in the LSST images is not ideal for space-based images without a large atmospheric PSF. To improve the visual realism of the images and avoid detection of these star-forming regions as unique stars, each knot is convolved with a Gaussian with $\sigma=0.2$ arcsec. This is about twice the PSF size of the Roman images, but still small enough that it should not cause substantial issues when comparing galaxy models with the LSST images, which have a median PSF FWHM of about 0.67 arcsec. Third, objects in the Roman HLIS images are rendered chromatically by integrating over the full bandpass and SED of the object instead of evaluating the flux at a representative wavelength.\footnote{Some chromatic effects like differential chromatic refraction were included in the LSST image simulation.}

\section{Wide-Field Survey Strategies}\label{surveys}

The synthetic wide-field surveys produced for the Roman HLIS and Rubin LSST are constrained by the characteristics of each observatory. We summarize these limitations and their impact on the simulated survey sequences used to create these joint simulations below.

\subsection{Roman High-Latitude Survey}\label{surveys-roman}

The Roman High-Latitude Survey consists of both imaging and spectroscopic parts. The Reference Survey concept, used to define requirements on the observatory and ground system, is used in our simulations. We emphasize that the actual survey executed by Roman may be significantly different: for example, the depth vs.\ area trade could be re-optimized, or one could do a multi-tier survey and re-allocate some time to a wide layer \citep{2021MNRAS.507.1514E}. Here we focus on the imaging (HLIS) component; the spectroscopic component (High-Latitude Spectroscopic Survey, or HLSS) is described in more detail in Sec.~2.2 of \citet{2022ApJ...928....1W}.

\begin{figure*}
\begin{center}
\includegraphics[width=\textwidth]{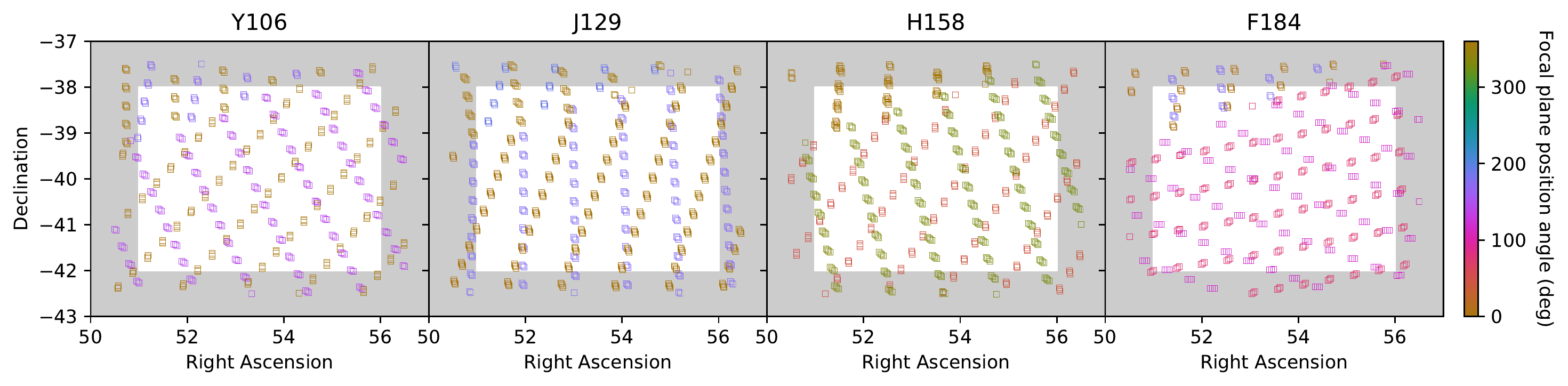}
\end{center}
\caption[]{The Roman Reference Survey HLIS dither strategy in the simulated region of the sky. The pointing coverage is shown in the survey cutout region (non-shaded area) in each of the four filters. The survey area is scanned along lines with four short dithers to cover chip gaps followed by a larger dither the size of the focal plane (described in Fig.~\ref{fig:imaging_tiling}). The telescope then scans back over the same region with a different filter or position angle.
\label{fig:pointings}}
\end{figure*}

The Reference HLIS is a 4-filter survey, with the filters Y106, J129, H158, and F184 covering the wavelength range from 0.92 to 2.00 $\mu$m, with typically 5--6 dither positions per filter after accounting for loss of coverage due to chip gaps and 140 s per exposure. This survey mode reaches a $5\sigma$ point source depth of $\sim 26.9$ mag AB in Y106, J129, and H158, and 26.2 mag AB in F184 \citep{2019arXiv190205569A}. The Reference HLIS encompasses 1 year of observations, interspersed with other programs during Roman's 5-year primary mission; this allows coverage of 2000 deg$^2$ plus deep fields (not simulated here). The Roman HLIS footprint is placed within the LSST baseline WFD footprint to enable joint analyses. Details of the HLIS survey design and tiling can be found in Appendix~\ref{hlisdef}. We show the pointings and their angles that overlap the 20 deg$^2$ simulated in this paper in Fig.~\ref{fig:pointings}.

\subsection{LSST Wide-Fast-Deep Survey\label{sec:rubinwfd}}

While the details of the simulation used in DC2 can be found in Sec.~3 in \citetalias{2021ApJS..253...31L}, we briefly summarize them here. We use an early baseline operational simulation (minion\_1016)\footnote{\url{https://docushare.lsst.org/docushare/dsweb/View/Collection-4604}} produced using the LSST Operations Simulator (OpSim; \citealt{delgado+2014}; now replaced by the scheduler introduced in \citealt{2019AJ....157..151N}); the simulation output includes a baseline WFD and five DDFs. Since OpSim did not inherently include dithers (translational or rotational) of the 3.5$^{\circ}$ LSST field-of-view (FOV), while it tiled the sky with a static grid of equal-area HEALPix pixels and pointed the FOV at the centers of the hexagonal pixels, the pointings were post-processed to include random, large\footnote{Large here means on the scale of the FOV.} translational dithers every night for survey uniformity (following the findings of \citealt{awan+2016}) and random rotational dithers after every filter change (which helps with shear systematics, as investigated in \citealt{almoubayyed+2020}); note that the current baseline simulations now have these dithers built-in. The specific DC2 area was chosen to be representative of the WFD survey, avoiding low Galactic latitudes and yielding uniform depth coverage. Figs.~1--2 in \citetalias{2021ApJS..253...31L} show the DC2 footprint, while Table 3 in the paper includes the exact coordinates of the area.

\section{Simulating Roman HLIS and LSST images}\label{sec:imagesims}

Simulations of both the Roman HLIS and LSST images are based on GalSim\footnote{\url{https://github.com/GalSim-developers/GalSim}} \citep{rowe15}. The Roman HLIS image simulation pipeline is described in \cite{2021MNRAS.501.2044T}, while the LSST image simulation pipeline is described in \citetalias{2021ApJS..253...31L} and \citetalias{2021arXiv210104855L}. The LSST images used in this work are unchanged from these references.

\subsection{The Roman HLIS image simulation pipeline}

The Roman HLIS image simulation suite has undergone several updates in how the telescope is simulated since \cite{2021MNRAS.501.2044T,2022arXiv220308845Y,2022arXiv220413553W} to improve realism in the physics that occurs within the optics and detector systems. These are described in the paragraphs below. In addition, the simulation pipeline now has more sophisticated stages to post-process the imaging data to perform basic coaddition, background subtraction, detection, and photometric measurements. These new post-processing steps are described in Sec.~\ref{results}.

\textbf{PSF:} The PSF is no longer simulated using an approximation of the Roman pupil plane as it was in \cite{2021MNRAS.501.2044T}. Instead, depending on the needed accuracy, a high-resolution (2048$\times$2048) filter-dependent\footnote{Roman has an exit pupil mask on each filter in the element wheel. For the reddest filters, the mask is oversized to block thermal emission from the warm telescope, including the spider.} image of the pupil plane for each SCA is binned and used to simulate the PSF. The pupil image matches Roman Cycle 7 specifications and is binned every 8 pixels to calculate the PSF to be convolved with all galaxies and any stars fainter than magnitude 15. This includes all stars that might be usable for PSF modeling. For brighter stars we use a binning of either 2 or 4 pixels to prevent visual artifacts in the rendered images. The most significant result of this change is that bright stars now have the appropriate number of diffraction spikes (12) due to the six struts supporting the secondary mirror.

\begin{figure}
\begin{center}
\includegraphics[width=0.5\columnwidth]{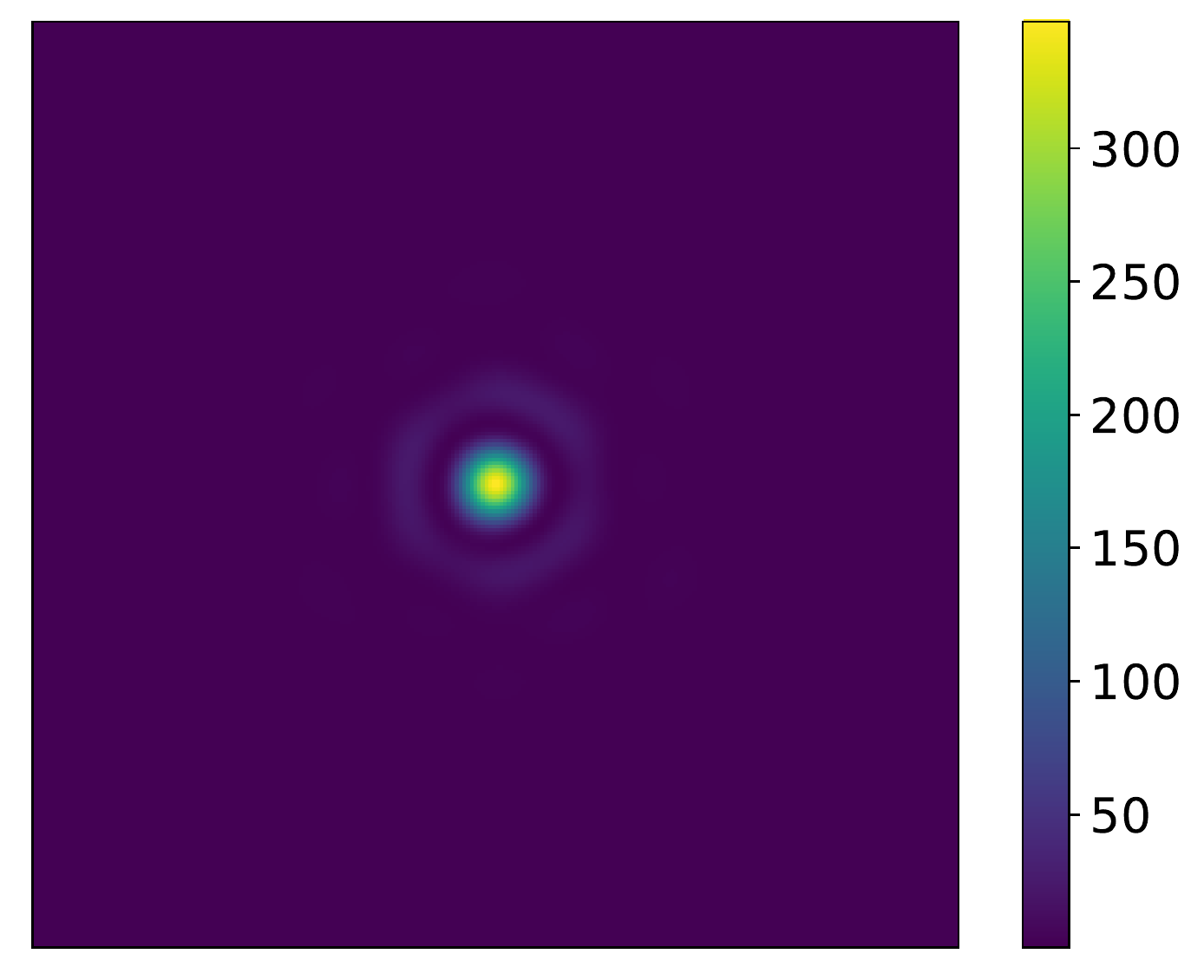}
\includegraphics[width=0.49\columnwidth]{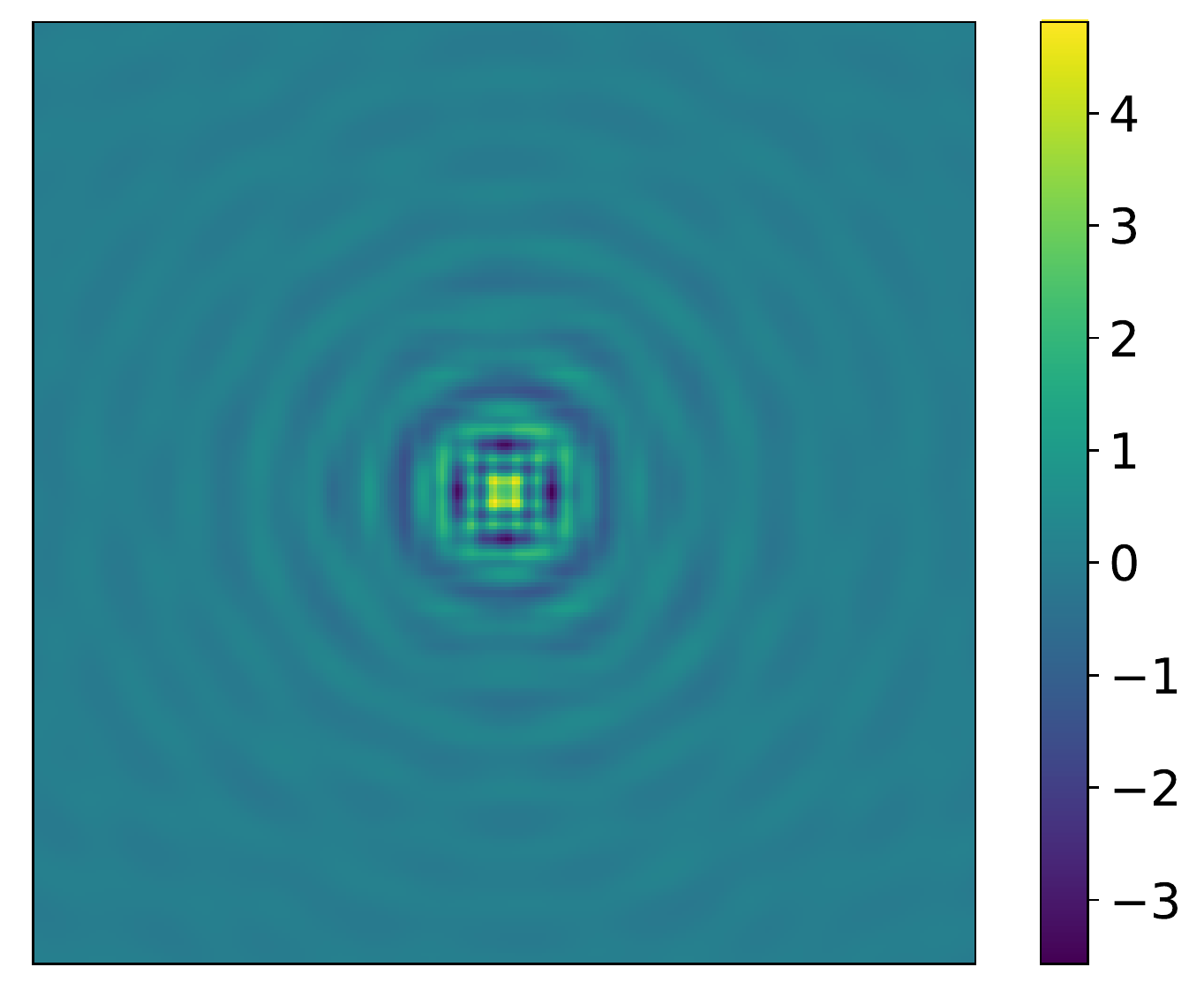}
\includegraphics[width=0.5\columnwidth]{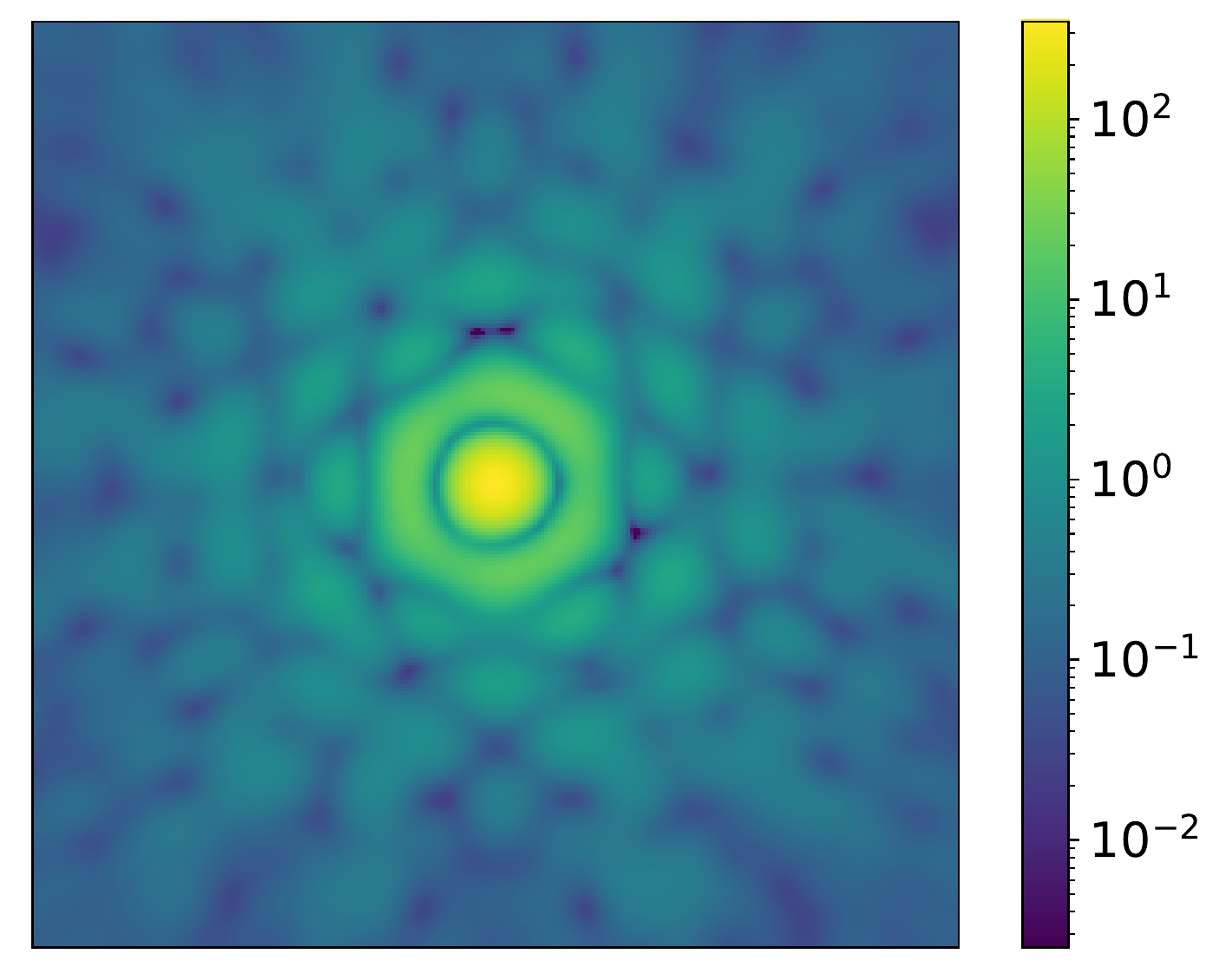}
\includegraphics[width=0.49\columnwidth]{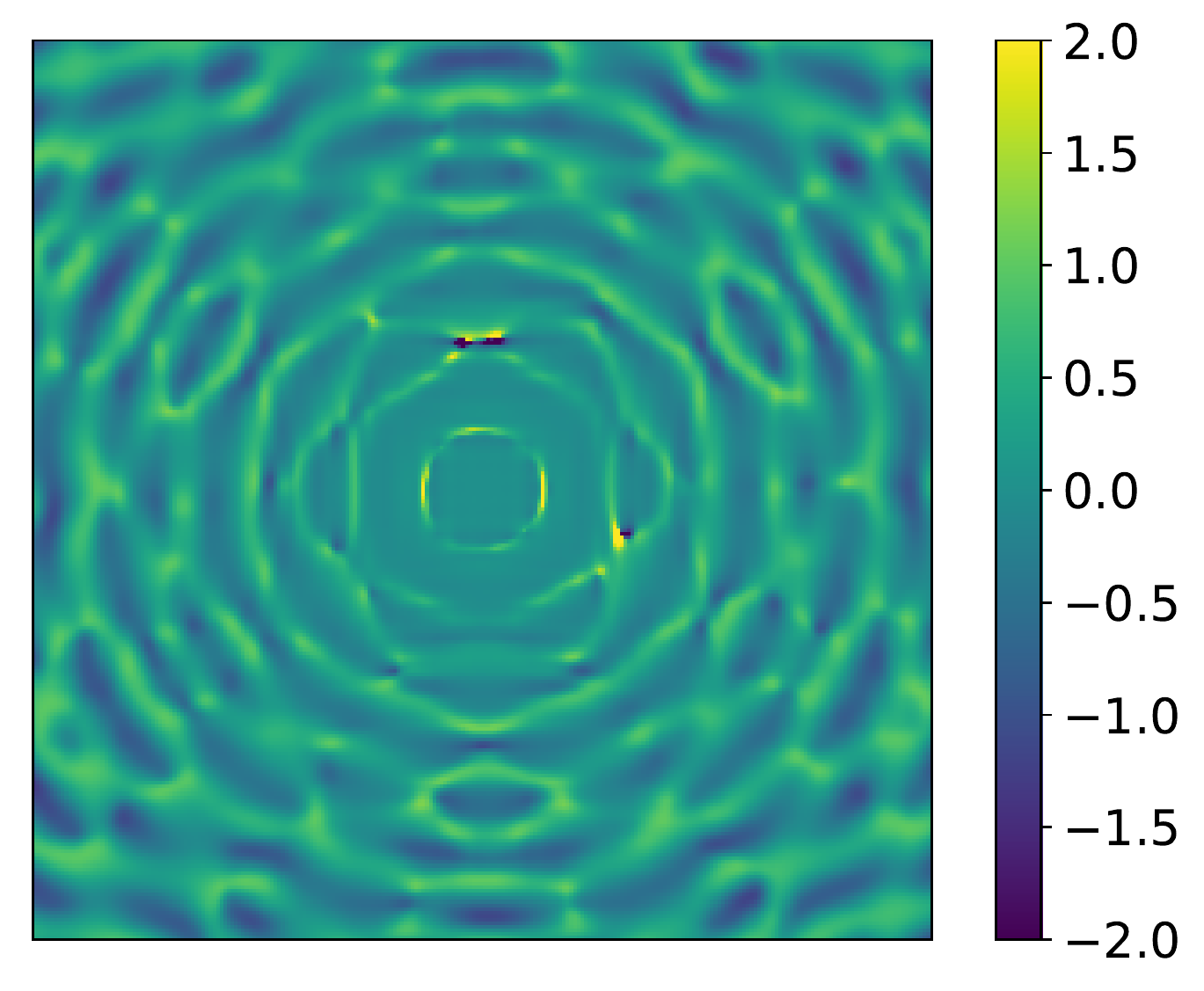}
\end{center}
\caption[]{The impact of chromatic vs.\ achromatic simulation of the Roman PSF. Each PSF cutout spans 1.375 arcsec (12.5 native Roman pixels).
Top left: The chromatic Roman PSF associated with the F184 filter and SCA 18 with linear scale.
Bottom left: The same PSF image with log scale.
Top right: The difference between the chromatic PSF image and the achromatic PSF image.
Bottom right: The fractional difference between the chromatic PSF image and the achromatic PSF image.
\label{fig:psf}}
\end{figure}

\textbf{Chromatic rendering:} This synthetic Roman survey is the first large-scale image simulation to use fully chromatic rendering of the PSF and object flux based on the objects' SEDs, which has been made computationally feasible with recent updates in GalSim.\footnote{Stars brighter than magnitude 15 are not chromatically rendered, but are unlikely to have any precision science use as they are saturated.} The LSST images were created before this update, and evaluate the PSF and flux at a fixed reference wavelength for each filter, like previous Roman simulations. Figure \ref{fig:psf} shows how this change impacts the Roman PSF.

\textbf{SCA physics:} The Roman SCAs can now be simulated with updated models for various physical effects in the detectors based on measurements of the flight detectors. A complete characterization of the Roman H4RG-10 detectors is provided by \cite{2020JATIS...6d6001M}. Models for many of these effects have been implemented in GalSim \citep{2016PASP..128i5001K,2016PASP..128j4001P,2022MNRAS.512.3312L}. We describe in more detail in Appendix~\ref{detectors} the physical processes we now simulate in the detectors with updates to this modeling. We have also allowed for spatial variations in SCA properties, and have built the first iteration  (``v1'') of the SCA properties based on laboratory measurements on the flight detectors using the solid-waffle code \citep{2020PASP..132a4501H,2020PASP..132a4502C,2020PASP..132g4504F,2022PASP..134a4001G}. The v1 files still contain some placeholders and there are some significant differences relative to the flight configuration (as detailed in Appendix~\ref{detectors}); but they support ongoing development of simulation and analysis tools, and serve as an initial test of the interface between calibration and simulation codes.

The simulation of objects in the Roman pipeline is physically equivalent to the process described below for the DC2 simulations, except for an issue with how dust extinction is applied. The Roman HLIS images have improperly applied the internal dust extinction of the simulated galaxy as Milky Way dust extinction. This means that the Milky Way dust extinction is not present in the Roman images or measurements, and instead, the rest-frame intrinsic dust model was applied in the observed-frame. This must be accounted for in measurements, and a dereddening correction that properly removes this effect is provided at the catalog level.

\subsection{The LSST DESC image simulation pipeline}

The LSST images were produced using the DESC image simulation package \textsc{imSim}, which utilizes the simulation utilities provided by GalSim.\footnote{\url{https://github.com/LSSTDESC/imSim}} A detailed discussion of \textsc{imSim}, and how it was used in the DC2 simulations can be found in \citetalias{2021ApJS..253...31L}. Here we briefly describe the version of \textsc{imSim} used for DC2 and the simulated images from the Rubin Observatory it produced.  The DC2 simulations were driven by a simulated LSST cadence described in Sec.~\ref{sec:rubinwfd}, which specified pointings, atmospheric conditions, filter choice, lunar position, etc.\ over a five year period.  With that information, an input file for \textsc{imSim}, known as an instance catalog, was generated for each exposure.  Each instance catalog contained overall information for each exposure such as telescope orientation and the choice of filter, along with  a list of sources to simulate.  The sources included both the CosmoDC2 galaxies and the Milky Way stars described in Sec.~\ref{dc2}.  For each galaxy the position, magnitude, shape, shear parameters, SED, and internal and Milky Way dust extinction parameters were specified for each object component. The lensing shear parameters were extracted for each galaxy via ray-tracing from the N-body `Outer Rim' simulation.

First, the photons from the sources were ray-traced through a multi-layer turbulent atmosphere to build an atmospheric PSF.  Then, the optics of the telescope were simulated with a parametric model including residual optical aberrations.  Finally, the photons that reach the focal plane were propagated into the silicon of the sensors.  A custom-built silicon sensor-model~\citep{2019arXiv191109577L} was used by GalSim to simulate diffusion and the brighter-fatter effect.  Other sensor effects such as tree-rings and bleeding were also implemented in \textsc{imSim} and used for DC2.

Next, in the readout stage, electrons were converted to ADUs via a gain with electronic noise. Cosmic rays were also added.  The resulting output included a file containing the positions and fluxes or the true sources, a FITS true-image before electronics readout, and a fully simulated FITS file containing each of the sixteen amplifiers for all simulated sensors in the focal plane.

The resulting fully simulated images and accompanying simulated calibration products were then processed by Rubin's LSST Science Pipelines\footnote{\url{pipelines.lsst.io; v19}} \citep{2019ASPC..523..521B} to produce calibrated and processed images and co-adds along with their associated source catalogs.

\section{Simulation results}\label{results}

The joint-survey Roman HLIS--LSST simulated imaging footprint is defined by a coordinate rectangle
$51<\textrm{RA}<56^\circ$ and $-42<\textrm{Dec}<-38^\circ$, covering an area of 20 deg$^2$. This footprint is shown in Fig.~\ref{fig:pointings}. There are 32,927,931 galaxies and 218,080
stars with $r<29$ simulated in this region. While the limiting magnitude of Roman HLIS and LSST images
are much brighter than this limit, we include these faint sources to properly simulate the
correlated sky background. We are able to detect in the Roman HLIS coadd images 12,432,713 galaxies and
215,613 stars with signal-to-noise greater than five in the median detection image, while the LSST pipeline detects 6,831,656 galaxies and 159,575 stars in the LSST coadd images. Direct comparison of the observed numbers of objects between the surveys should be made with caution, given the difference in maturity and development of the two pipelines used here. In particular, the Roman coadd, detection, and measurement pipeline is not what will be developed for use during the mission.

\subsection{Coadd images, detection, and photometry}
\label{sec:coadddet}

We split the Roman HLIS survey area into rectangular regions, within which we coadd individual exposures into coadd images. Each coadd image has a unique, non-overlapping area of 0.0156 deg$^2$. Each full coadd image is 8825$\times$8825 pixels, including a 500 pixel overlap on all edges with adjacent coadd tiles. The coadd pixel scale is 0.0575 arcsec, chosen so that coadd images in J129 are Nyquist sampled. The typical number of single-epoch images contributing to the coadd image in each filter at any location is 5-6. In total, there are 1039 coadd tiles overlapping the survey region footprint. Example coadd image cutouts are shown in Figs.~\ref{fig:diffim} \&~\ref{fig:imageexample} and example color cutouts built from the coadd images are shown in Fig.~\ref{fig:cutouts}.

Basic object detection and segmentation is performed on a median detection image built from Y106, J129, H158, and F184 coadd images using Source Extractor \citep{1996AAS..117..393B}, rather than taking the union of detections in each bandpass coadd. Basic photometric information is provided for each bandpass coadd, with a 5$\sigma$ model detection threshold based on the median image. The truth catalogs contain a dereddening correction in each bandpass, which can be subtracted from the true and measured magnitudes, and the object catalogs contain a quality flag that is the bitwise `or' of the Source Extractor flags from each of the detection and per-bandpass forced photometry measurements. We further recommend using a signal-to-noise greater-than-five selection in the median detection image flux. All magnitudes in this work have been calibrated by subtracting the mean difference in the measured and true magnitudes of stars between magnitudes 17 and 20 to mimic photometric calibration using standard stars.

We enforce several selection: 1) removing any objects flagged by Source Extractor, which removes 32\% of detections; 2) an explicit requirement that the objects have signal-to-noise greater than five in the detection image, which removes a further 0.03\% of objects; and 3) we only use objects that have a positional match to a true object throughout this work, which removes a further 1.7\% of objects. This is determined by finding a matching true object within 1 arcsec, and if there are multiple matching objects, taking the closest matching object in magnitude between the up to three nearest. Further details of the processing of the Roman HLIS images are in App.~\ref{app:coadd}.

\begin{figure*}
\begin{center}
\includegraphics[width=\textwidth]{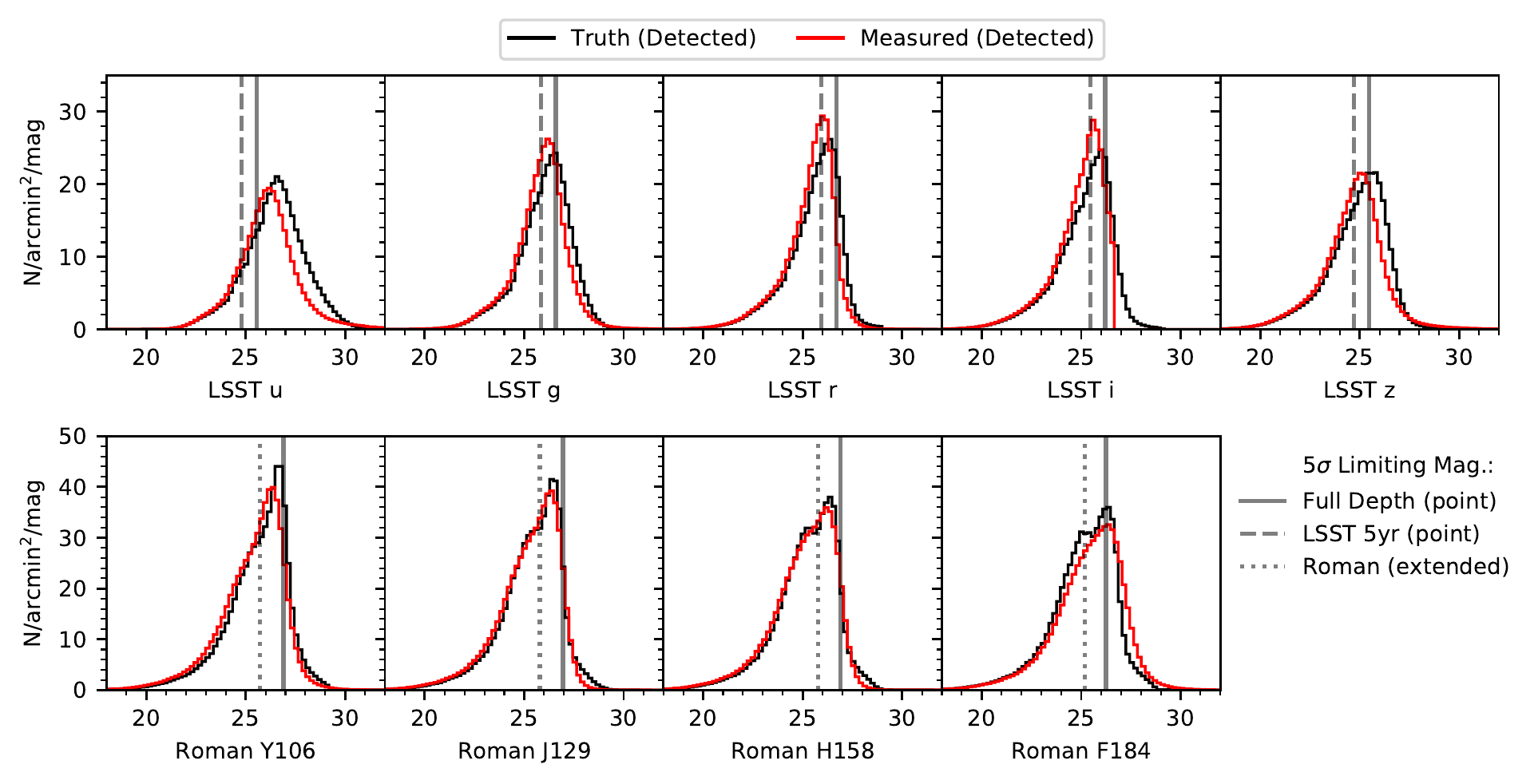}
\end{center}
\caption[]{A comparison of the measured (red) and true (black) magnitudes for objects detected in the simulation across bandpasses in the Roman HLIS and LSST surveys. We show several estimates of the expected limiting magnitude for the surveys as vertical lines. Solid lines indicate the 5$\sigma$ point-source limiting magnitude for full-survey depths. Dashed lines show the same limiting magnitude for five-year survey depth for LSST. Dotted lines show the limiting magnitude for a 0.3 arcsec extended source for Roman. There is generally goody agreement in the measured and expected depths, though we see a skew to fainter magnitudes in the measured F184 histogram that may have to do with the different noise properties in this bandpass due to infrared glow of the telescope.
\label{fig:galmag}}
\end{figure*}

\begin{figure*}
\begin{center}
\includegraphics[width=\textwidth]{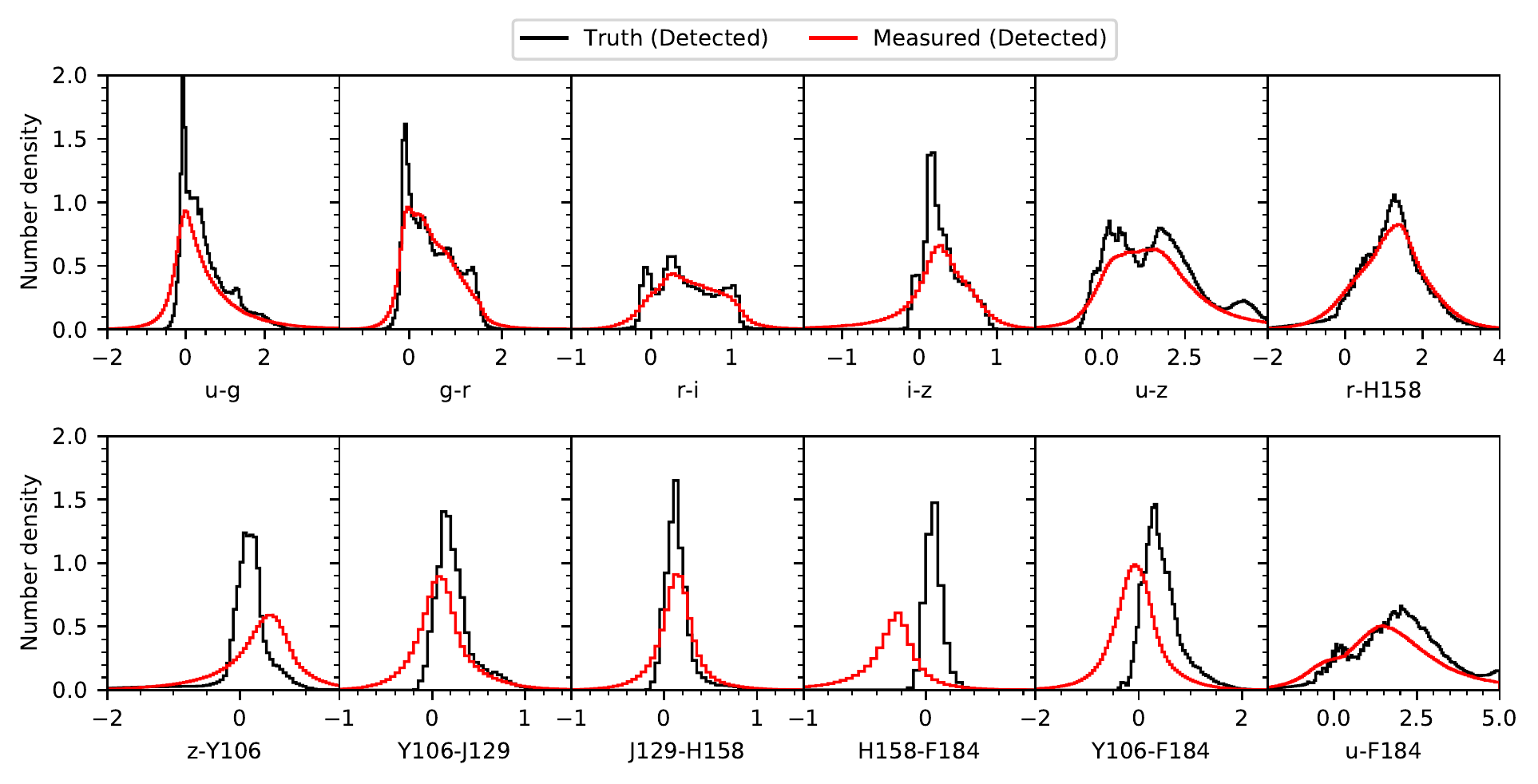}
\end{center}
\caption[]{A comparison of the measured (red) and true (black) colors for objects detected in the simulation across filters in the Roman HLIS and LSST surveys. Overall, the agreement is reasonable, but there appears to be in particular an issue with the measured colors with F184 in Roman relative to the truth. This is likely related to the observed skew to fainter measured magnitudes vs true in F184. A grey photometric correction based on comparison of measurements for bright, unsaturated stars to true magnitudes was performed, but this may indicate an additional chromatic correction to the measured photometry is needed.
\label{fig:galcol}}
\end{figure*}

The simulated LSST DC2 images are processed as described in Sec.~7
of \citetalias{2021ApJS..253...31L}.  In brief, the steps are as follows:
\begin{enumerate}
\item The LSST Science Pipelines were used to process
individual frames, including instrument signature removal, basic image characterization steps such as initial photometric and
astrometric calibration, background estimation, source detection and measurement, and PSF modeling, all carried out at the level of
individual CCDs.  This step produces Processed Visit Images (PVIs).
\item Then the images were resampled to a common pixel grid with $0.2\arcsec$ pixel scale, and coadded; the PSF models were
coadded using the same weights as the images to produce PSF models for the coadd.
\item After that point, sources were detected, deblended, and characterized on the coadds, including carrying
out forced photometry to obtain robust color measurements.  The resulting catalogs are referred to
as ``object catalogs.''
\end{enumerate}


\subsection{Data products}

The Roman HLIS simulation data products include a variety of catalogs and images. These include:
\begin{enumerate}
\item A FITS catalog describing the Reference HLIS observing sequence.
\item Separate FITS catalogs for both the stars and galaxies that contain the full set of truth information necessary to reproduce the image-level data, with a supplementary HDF5 file containing all the model SEDs referenced for the stars and galaxies. These initial truth catalogs are a superset of the objects that end up in the simulated region of the sky, and the brightness of each component of the objects is calibrated using a 1~Angstrom wide function bandpass at 0.5~$\mu$m used in the DC2 simulations, called ``mag\_norm.''
\item Summary truth FITS catalogs that include the original object index to link to the catalogs in (ii) as well as several derived properties like the total true object model magnitude and where the object is simulated in the image. These files are provided per-SCA image and per-coadd image.
\item A set of `true' images for each SCA in the observing sequence that overlaps the targeted region of the sky, which include the Roman PSF and WCS, but no other instrumental effects or noise.
\item A FITS-based MEDS-like library\footnote{\url{https://github.com/esheldon/meds}} of cutouts of each object from the `true' images, including their local WCS.
\item A set of observed images, which include (i) 29,153 images using a `simple model' of just the appropriate noise (including dark current) and saturation, and (ii) a subset region of 6653 images using the full `SCA model' using the detector physics described in App.~\ref{detectors}.
\item A FITS image file with a high-resolution PSF image for each SCA in each pointing, oversampled relative to the native Roman pixel scale by a factor of eight.
\item A set of 1039 coadd images in each of the four (Y106/J129/H158/F184) bandpasses based on the `simple model' images.
\item A FITS image file with a high-resolution PSF image for each unique combination of SCA images for each coadd image, oversampled relative to the coadd pixel scale by a factor of eight. A lookup function is provided to retrieve and downsample this PSF cutout image to a given coadd stamp size and pixel scale.
\item A set of FITS image segmentation maps and FITS detection and photometry catalogs for each set of four coadd images.
\end{enumerate}

Details on the DC2 simulated LSST data can be found in section 8 of \citetalias{2021ApJS..253...31L} and
in \citetalias{2021arXiv210104855L} (which covers only the subset that is made fully public), and
like the Roman data products, include a variety of truth, image, and catalog data.  Here we mention the key products
that were used for this work, rather than summarize all of them, and refer interested readers to
the referenced works for more comprehensive information.
\begin{enumerate}
\item Truth catalogs contain the information about the simulated objects, which for galaxies and
static stars can be represented as a single table without any time variability, and for time-varying
objects (variable stars, supernovae, etc.) is more complex.  Truth catalogs include
information about position on the sky, redshift, type of object (point source or extended, static or
variable), and flux in each band without and with Milky Way extinction.  For variable objects, there
is additionally a table providing information about each observation, representing the change in
flux from the summary table.
\item Object catalogs contain information about the detected sources on the coadd, after merging
detections across bands and carrying out forced photometry with respect to some reference band.
These include a number of quality flags, information about the deblending process, a flag for
extendedness (to separate point sources and extended objects), position and color information, and
more.
\item In some cases we also used the coadded images, for example to compare images in Roman HLIS  and
DC2 simulations in some region.
\end{enumerate}


\begin{figure}
\begin{center}
\includegraphics[width=\columnwidth]{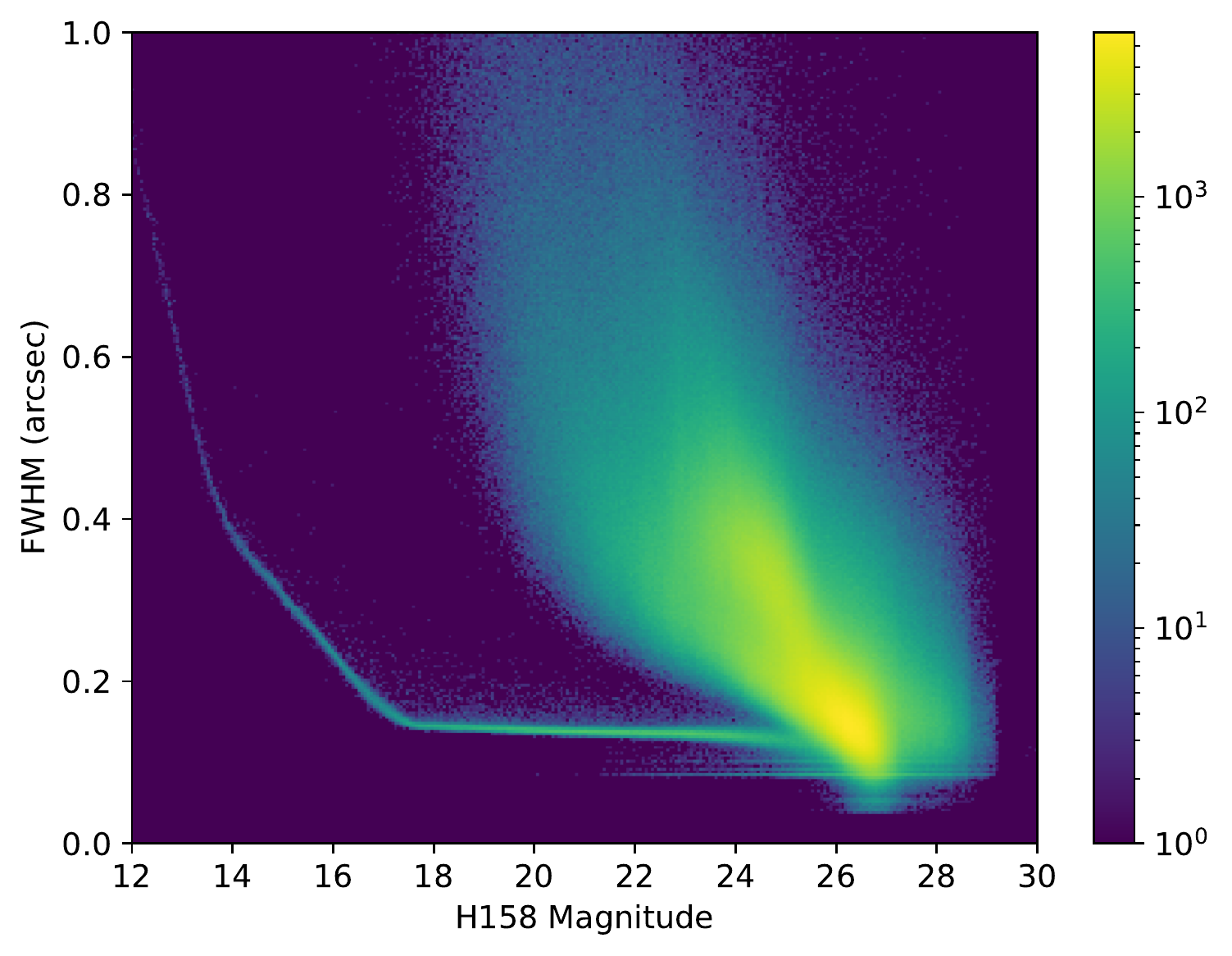}
\end{center}
\caption[]{The distribution of measured size vs.\ H158 magnitude from the Roman HLIS detection catalog. The stellar population is clearly distinguishable in the lower branch. The impact of saturation is apparent at magnitudes brighter than approximately 17.
\label{fig:locus}}
\end{figure}

\begin{figure}
 \begin{center}
 \includegraphics[width=\columnwidth]{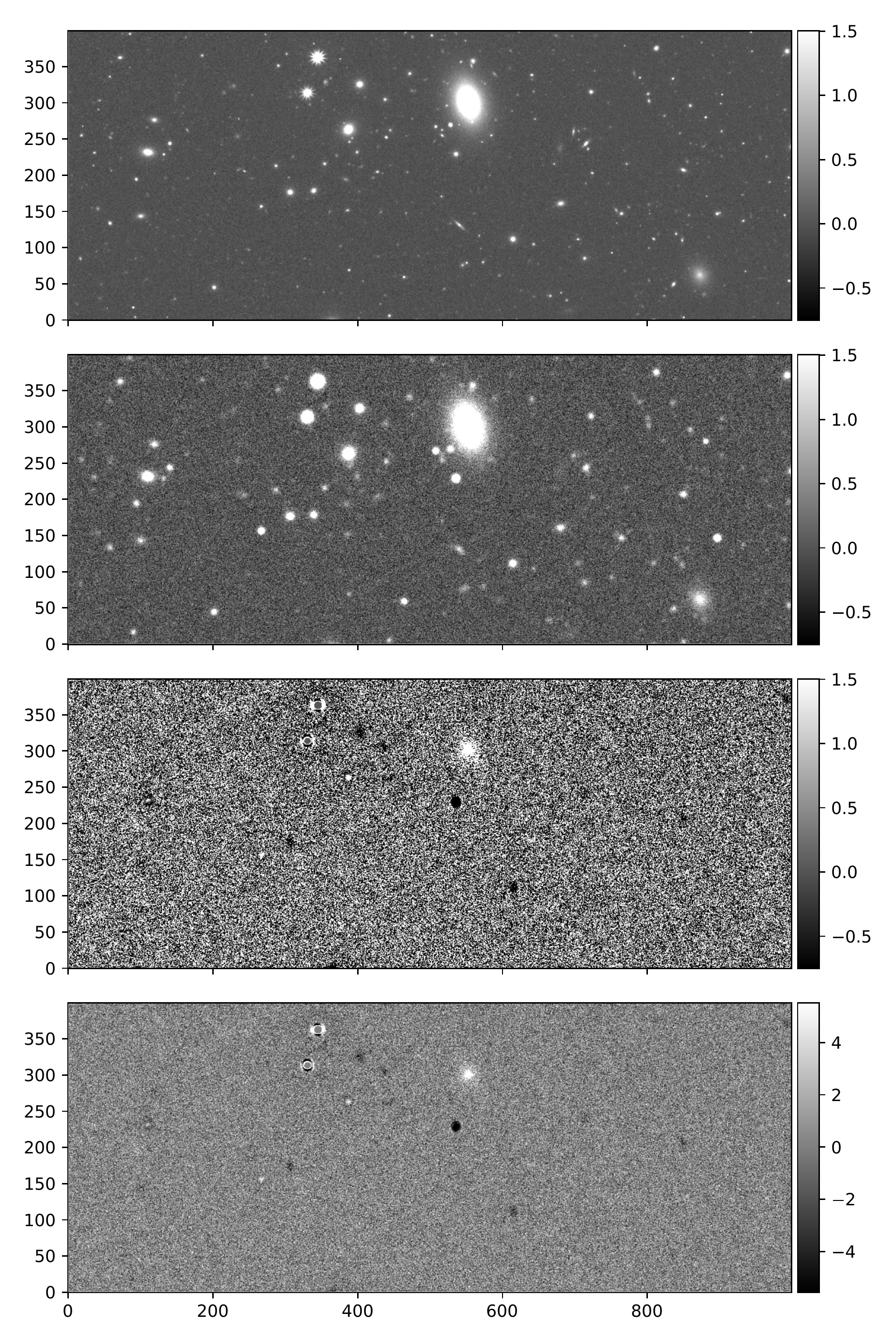}
 \end{center}
 \caption[]{Comparison of Roman HLIS and LSST simulated coadd images of the same field. Top to bottom: 1) Roman Y106-band coadd image, scaled to match LSST image in (2); 2) LSST $y$-band coadd image; 3) difference image calculated with HOTPANTS using the Roman HLIS image as a deep template for subtraction, scaled to match the LSST image in (2); 4) same difference image using a standard ZScale and linear stretch for better visualization of dynamic range. This comparison provides a general validation test of the two simulation pipelines against each other,  demonstrating qualitatively that the simulated universe and its physics are being simulated in consistent ways in each synthetic survey.
 \label{fig:diffim}}
 \end{figure}

\subsection{Validation of simulation output}

We validate the realism of the simulation for the Roman HLIS in the near-infrared and the data products and measurements produced in several ways. First, in Fig.~\ref{fig:galmag}, we compare the measured magnitudes for galaxies in Roman and Rubin bandpasses to the true photometry of objects detected in either synthetic survey. We also plot the predicted 5$\sigma$ limiting magnitudes in each survey and filter as vertical lines. The 10-year point-source prediction (solid) for LSST uses the LSST scheduler \citep{2019AJ....157..151N} and associated simulated pointing histories (‘Opsim outputs’) \citep{2016SPIE.9910E..13D,2014SPIE.9150E..14C} -- for this specific comparison, we use Opsim Run \textsc{baseline\_v2.1\_10yrs}   from the optimization process described in \citet{2022ApJS..258....1B}. The five-year dashed line is calculated by accounting for a factor of two in integrated exposure time. For the Roman HLIS, we use standard values for the Reference HLIS simulated in this work for both point sources (solid) and extended objects (dotted), with an exponential profile and half-light-radius of 0.3 arcsec\footnote{\url{https://roman.gsfc.nasa.gov/science/WFI_technical.html}} \citep{2013ascl.soft11012H}.
We find generally good agreement between the input and measured values and the expectations of the 5$\sigma$ limiting magnitudes of each survey.

We show the same comparisons for a variety of optical, near-infrared, and optical--near-infrared colors in Fig.~\ref{fig:galcol}. There is more variation here between the true and measured properties, though there is good agreement overall. The most significant difference is between the true and measured colors  that include the Roman F184 bandpass. While we applied a grey correction to the measured Roman HLIS photometry based on a comparison of the measured magnitudes of bright, unsaturated stars to their true magnitudes, we have not explored any color-dependent corrections for this work. Quantitative studies of, e.g., photo-$z$s using these simulations may require further work in calibrating the measured photometry or more a sophisticated measurement process than used here. Finally, we also demonstrate the stellar magnitude--size relation present in the near-infrared measurements using the synthetic Roman HLIS data in Fig.~\ref{fig:locus}.

As a general test of the two simulation pipelines and qualitative check that the simulated universe and its physics are being simulated in the same way in each synthetic survey, we compare in Fig.~\ref{fig:diffim} a Roman HLIS Y106- and LSST $y$-band coadd image. To highlight any differences, particularly in where and how object components are simulated in the images, we calculate a difference image using the HOTPANTS algorithm \citep{1998ApJ...503..325A,2015ascl.soft04004B}. We find that the difference comparison demonstrates there are no relative registration or astrometric issues between the two synthetic surveys, which would have resulted in a dipole subtraction. Instead, except for some minor artifacts from the subtraction process, the resulting differences are consistent with differences in the Roman Y106 and Rubin $y$ bandpasses. In almost all cases, though, no residual flux from objects is visible at the level of the noise in the difference operation.

\subsection{Analysis of blending in the Roman HLIS and LSST}

\begin{figure*}
 \begin{center}
 \includegraphics[width=\textwidth]{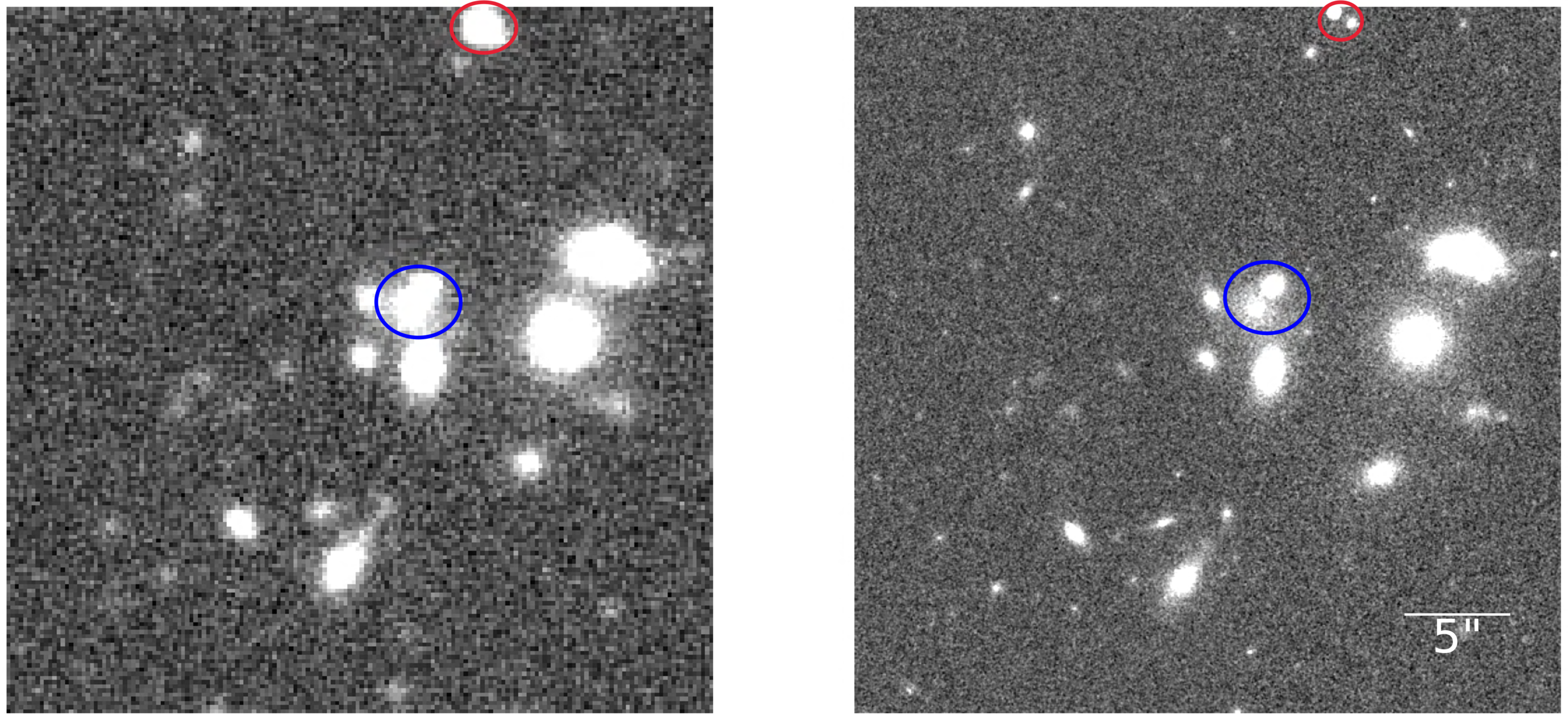}
 \end{center}
 \caption[]{A comparison of an overlapping region from LSST $i$-band (left) and Roman H158-band (right) coadd images. The regions circled in red or blue demonstrate a grouping of at least two objects that are blended in the LSST image, but clearly distinguishable as multiple objects in the Roman HLIS image.
 \label{fig:imageexample}}
 \end{figure*}

\begin{figure}
 \begin{center}
 \includegraphics[width=\columnwidth]{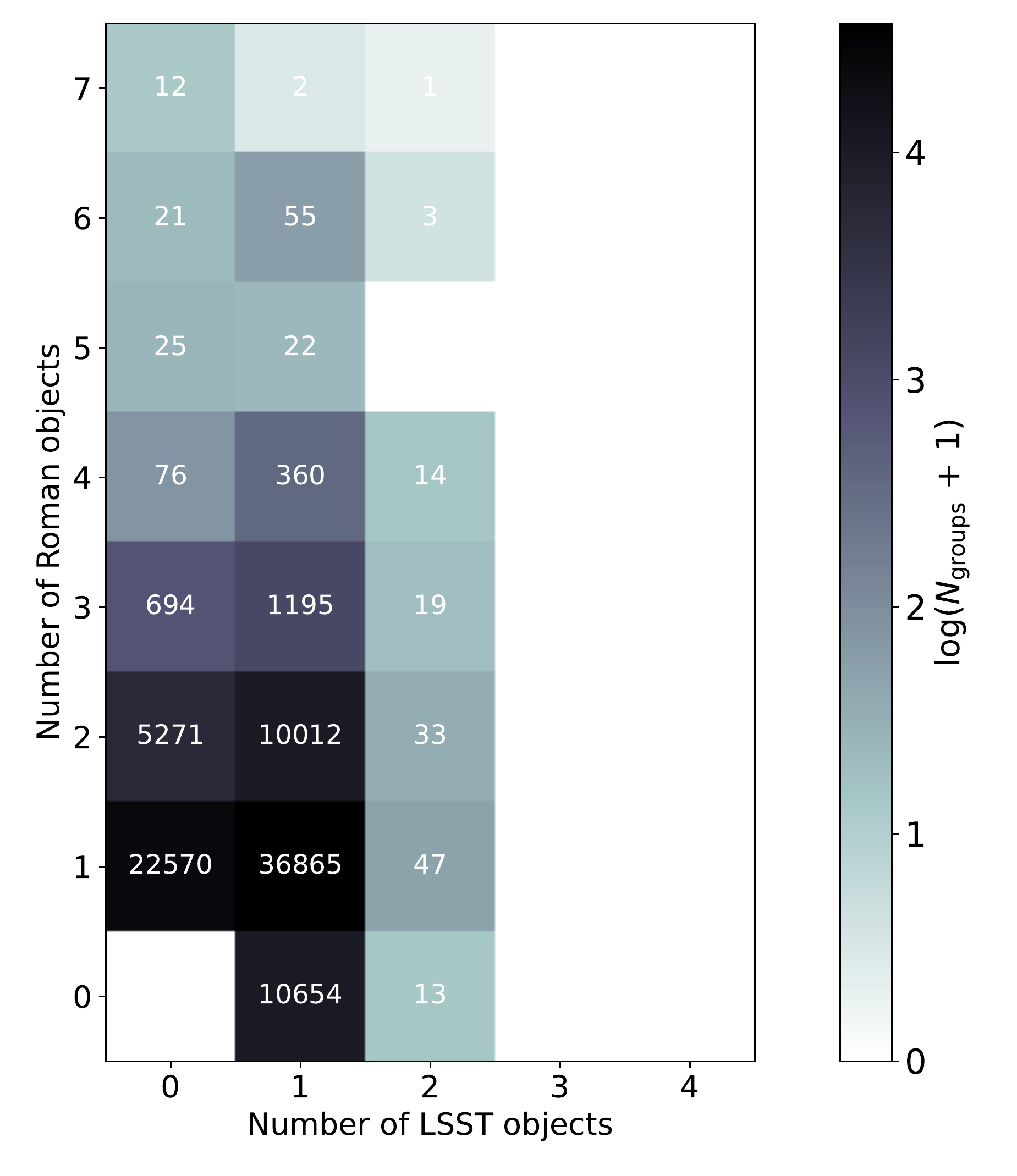}
 \end{center}
 \caption[]{Result of cross-matching a set of clean detections from LSST and the Roman HLIS. The numbers in each square
 correspond to the number of systems that have the corresponding number of friends-of-friends matching systems in the Roman HLIS (rows) and LSST (columns) images. As expected, we find more objects detected in the Roman HLIS than LSST and, for example in the second column, many systems of objects detected as a single LSST object are identified as multiple objects in the Roman HLIS images.
 \label{fig:fof}}
 \end{figure}

 \begin{figure}
 \begin{center}
 \includegraphics[width=\columnwidth]{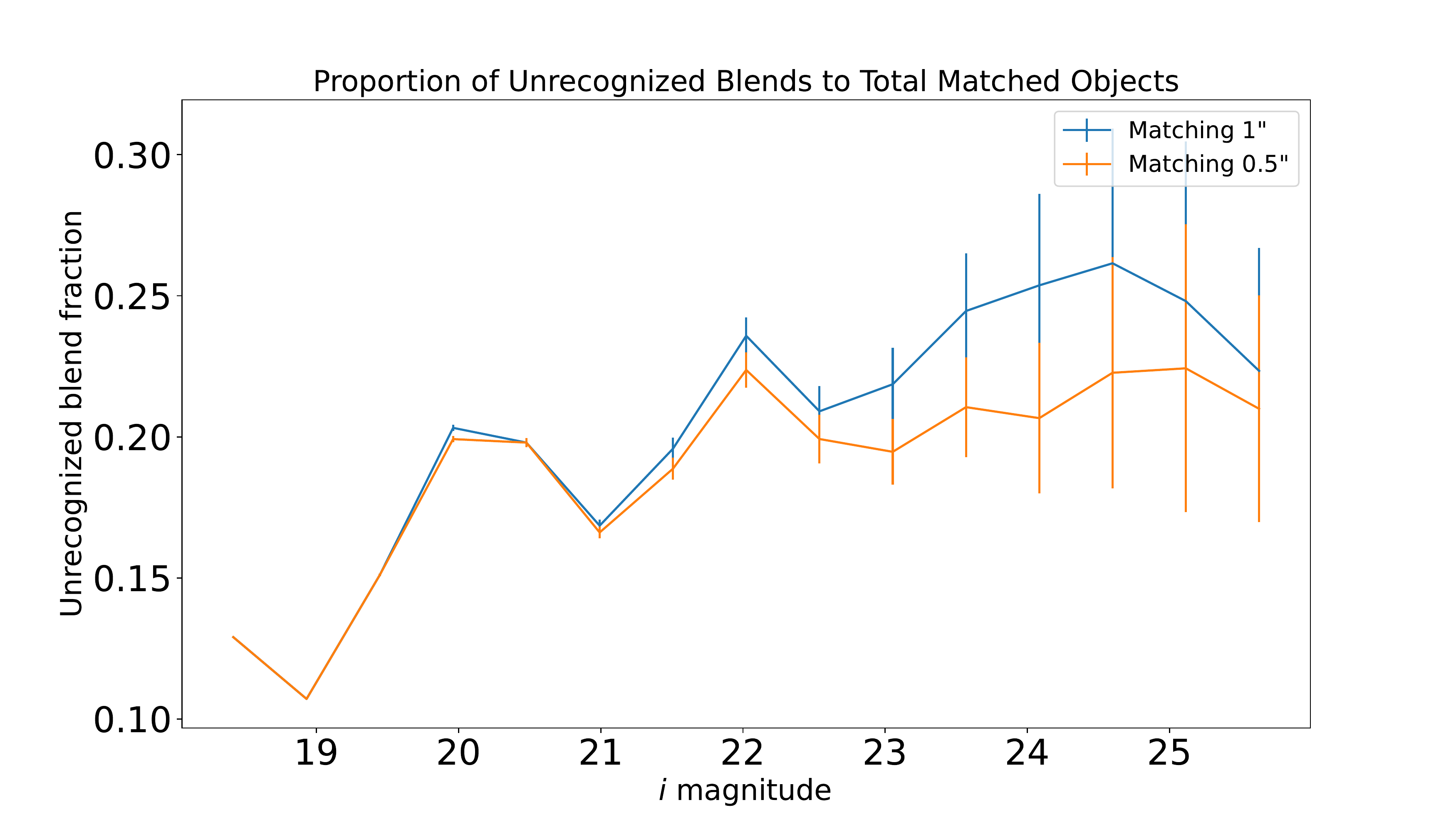}
 \end{center}
 \caption[]{Proportion of unrecognized blends to the total matched objects against the $i$-band
 magnitude of objects. The blue (orange) curve shows the rate of unrecognized blends as a function of
 magnitude determined with a matching radius of 1 (0.5) arcsec.
 \label{fig:magdist}}
 \end{figure}

 In ground-based surveys, blending between galaxies occurs more frequently than in space-based
 surveys due to the larger PSF caused by atmospheric turbulence \citep[e.g., see direct study
 in][]{2016ApJ...816...11D}. Failure to properly deblend object light profiles can affect the flux
 and second-moment measurements, thus impacting photometric redshift and shear
 estimation \citep[e.g.,][]{2021JCAP...07..043S,2022MNRAS.tmp.1305N}.  By
 looking at the overlapping area between the two surveys, the space-based images can be used to
 identify unrecognized blends and improve the deblending algorithm used for the ground-based data, or can be used as inputs to
 improve joint photometric redshifts \citep[e.g.,][]{2021MNRAS.500..531A}. In this section we
 present a preliminary study of the rate of {\em unrecognized blends} in five-years of LSST imaging that could be
 identified as such in Roman HLIS imaging, i.e., objects detected
 in LSST that are actually composed of more than one detected object in Roman HLIS imaging. We show an example of this in Fig.~\ref{fig:imageexample}, where blends of at least two objects in the LSST image are clearly distinguishable as multiple objects in the Roman HLIS images.

 For this study, we must begin with a complete set of detections that minimize false detections in the Roman HLIS and LSST simulated
 images following coaddition.  To obtain these, we carry out the following analysis steps following the selections described in Sec.~\ref{sec:coadddet}:
 \begin{itemize}
 \item Remove objects that have flagged pixels in the LSST detection catalog, but include objects with flagged value of 1 and 2 from the Source Extractor for Roman HLIS catalog to keep objects that are biased by neighboring sources and that have been deblended.
 \item Remove the extinction due to Milky Way dust from the measured LSST magnitudes using \textsc{dustmap}\footnote{\url{https://github.com/gregreen/dustmaps}} \citep{2018JOSS....3..695G} and from the measured Roman magnitudes using the provided extinction correction.
 \item Select galaxies in the LSST and Roman HLIS catalogs that have an entry in their respective
    truth catalog. This is done by cross-matching the object catalog to their truth catalog to
    remove spurious detections. 
     Cross-matching catalogs requires finding pairs of detections in the object and truth
    catalogs that correspond to the same object, by spatially joining the catalogs based on objects'
    positions. Objects less than some arbitrary distance apart would signify a match. We
    use \textsc{FoFCatalogMatching}\footnote{\url{https://github.com/yymao/FoFCatalogMatching}} \citep{2021ApJ...907...85M},
    which uses the friends-of-friends method to iteratively match multiple catalogs without specifying the main catalog.
 \item Apply the following magnitude cuts: $i < 25.63$ and $\textrm{H158} <
    25.42$. These values are obtained by fitting a power law over to each magnitude distribution over $[21, 26]$ and finding when the distribution falls below $80\%$ of the fit, which gives us a consistently defined `completeness' cutoff in each survey.  At fainter magnitudes, the
    catalogs are significantly incomplete, which complicates the analysis, and we do not attempt to model partial incompleteness at magnitudes between the limiting magnitude and this cutoff.
 \end{itemize}

 To identify separate Roman HLIS objects among blends in the LSST images, we cross-match the resulting
 catalogs of clean detections against each other.   The output of each FoF group can have multiple
 detected objects from multiple surveys.

 We first perform the analysis using friends-of-friends matching radius of 1.0~arcsec and lower it to
 0.5~arcsec to study the effect of matching radius. Figure~\ref{fig:fof} shows the result of
 cross-matching the two catalogs using a matching radius of 1.0 arcsec. The numbers in each square
 correspond to the number of systems that have the corresponding number of detected objects in the Roman HLIS (rows) and LSST (columns) images. We expect to see more instances of objects
 detected in the Roman HLIS but not LSST rather than the reverse, since the full 5 years of the Roman HLIS
 is expected to be deeper than the first five years of LSST. Possible unrecognized blends are located
 in the second column (one LSST detection, but multiple Roman HLIS detections). 
 
 We find 12\% of all groups of objects have an LSST detection and zero
associated Roman HLIS detections, with $\sim95\%$ of these objects having magnitudes within 1 magnitude of the LSST magnitude cut. These objects tend to preferentially be brighter in the wavelengths probed by LSST vs those probed by Roman, which we quantify using the $g-$F184 color. The mean shift in $g-$F184 color for this sample of objects relative to the full sample is 0.7, toward a bluer color. This makes sense, given that these are objects LSST is preferentially detecting over Roman. We also find that their sizes tend to be larger than the full population.

Figure~\ref{fig:magdist} shows the
unrecognized blend fraction for the two different matching radiuses used in cross-matching: 1.0 and
 0.5 arcsec. There are more unrecognized blends, more objects in one LSST detection and more than one Roman HLIS detections in Figure~\ref{fig:magdist}, with the 1 arcsec matching radius, because expanding the
 matching radius may convert some 1:1 LSST:Roman HLIS matches to be 1:2 or higher. For the 1 arcsec matching, there
 is a downward trend towards the dimmer end. This finding has several possible explanations: for example,
 two objects that are blended would appear brighter as a blend, shifting those objects to the left
 side of the distribution, or there are galaxies in the LSST detection catalog that are unrecognized blends, but one or both of the two galaxies are not detected in full-depth Roman HLIS images. The near-plateau for 0.5 arcsec matching likely implies a constant density of random chance projections, which is some function of the Roman-detected sky density.

 Using these same-sky LSST and Roman HLIS simulations, we find that typically $20-30\%$ of LSST detections at
 five-year survey depth may be unrecognized blends that could be detected as such using
 the Roman HLIS. \cite{2016ApJ...816...11D} studied the rate of unrecognized blends, or ambiguous blends, between Subaru SuprimeCam and HST, and identified that $18\%$ of the total number of Subaru objects are part of ambiguous blends. The rate of unrecognized blends between the DC2 and Roman HLIS simulations is broadly consistent with their finding. These studies of blending can potentially be used to diagnose and correct the effect of blending in LSST.  The fact that the
 simulated samples exhibit these unrecognized blending effects means that the simulations can be used
 in the future for tests of unrecognized blending and the impact on weak lensing, along with mitigation
 methods relying on joint analysis of both surveys.

 \begin{figure}
 \begin{center}
 \includegraphics[width=\columnwidth]{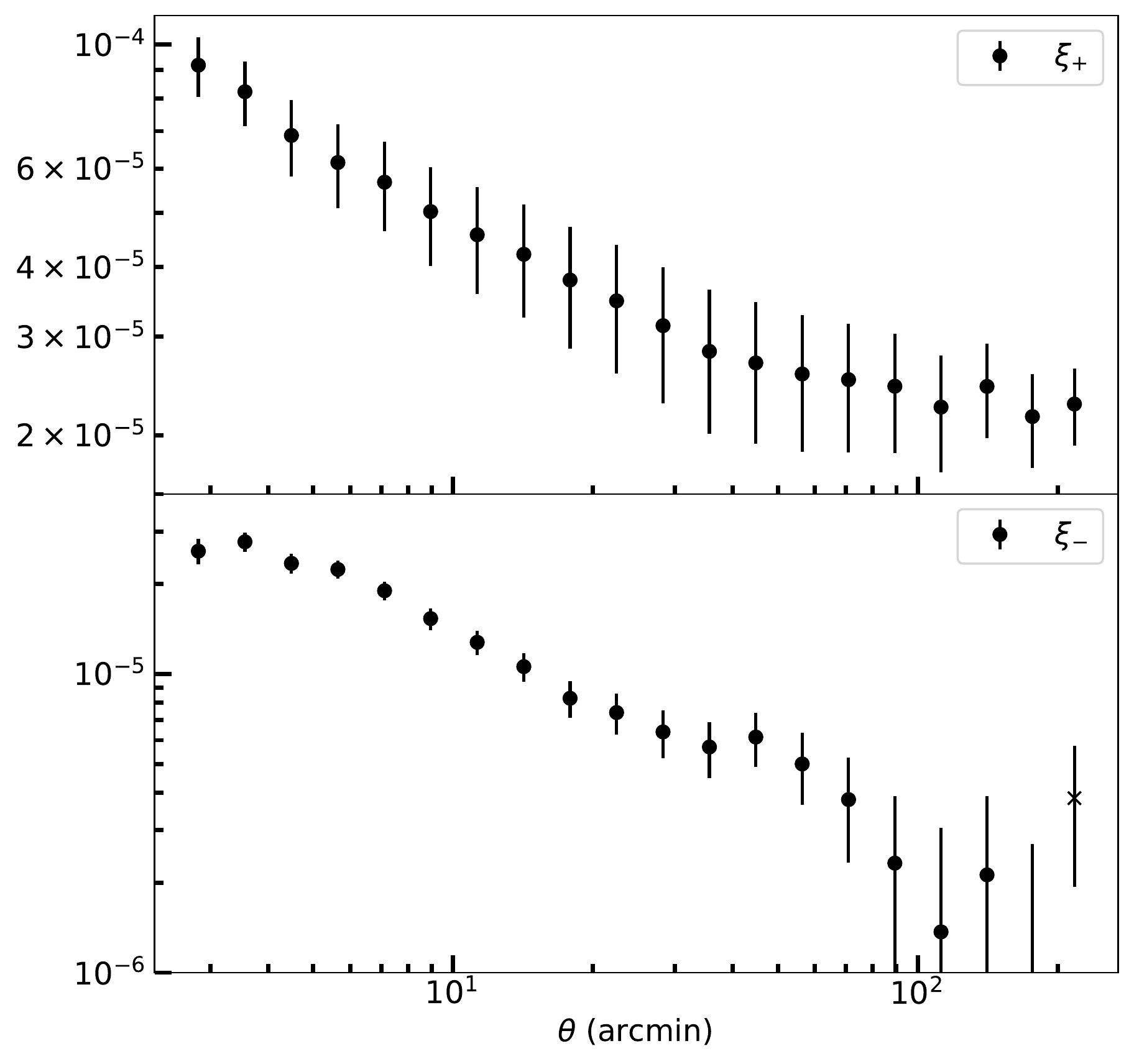}
 \end{center}
 \caption[]{The measured shear correlation function $\xi_{+}$ (top panel) and $\xi_{-}$ (bottom panel) for the Roman HLIS detection catalog using the true shape information. The error bars are computed from the square root of the diagonal of an analytic covariance matrix.
 \label{fig:xipm}}
 \end{figure}

\subsection{Statistical power of joint simulation region}

The joint synthetic survey area described in this work only covers 20~deg$^2$ of the sky, which is much smaller than the areas of the major ongoing weak lensing surveys like DES, HSC, and KiDS, which have wide-area observing fields in excess of 1,000 deg$^2$.  However, the extremely deep nature of the Rubin LSST (even if only simulated over the first half of its planned observing period) and the Roman HLIS surveys enables us to still measure cosmologically relevant weak lensing statistics to a precision similar to major ongoing weak lensing experiments.

We show the measured shear correlation function $\xi_{\pm}(\theta)$ for the objects we detect in the Roman HLIS coadd images using the true shape information in Fig.~\ref{fig:xipm}, with $\xi_{+}(\theta)$ in the top panel and $\xi_{-}(\theta)$ in the bottom panel. This is computed using {\sc TreeCorr}\footnote{https://github.com/rmjarvis/TreeCorr} \citep{treecorr} for a single (non-tomographic) redshift bin and 20 logarithmically-spaced angular bins from 2.5 to 250 arcmin. We use {\sc CosmoCov}\footnote{https://github.com/CosmoLike/CosmoCov}  \citep{2017MNRAS.470.2100K,2020JCAP...05..010F,2020MNRAS.497.2699F} to compute an analytic Gaussian covariance matrix to estimate the uncertainty, using the true redshift distribution and a galaxy density of 69 arcmin$^{-2}$ (all objects detected with detection-image auto flux signal-to-noise above 18) and a total $\sigma_e=0.35$. The correlation function signal-to-noise, estimated as $\sqrt{\xi_{\pm}^{T}C^{-1}\xi_{\pm}}$, is 47.

While this is just a simple example of a cosmologically relevant analysis that could be done using the synthetic Roman data set developed for this work, we emphasize that a wide range of wide-field science applications could be studied individually or jointly in the synthetic Roman HLIS and Rubin LSST surveys produced from the CosmoDC2 universe. This includes at the catalog level, as in this case to e.g. study combining observables in ground- and space-based data, but also jointly at the pixel level to e.g. perform joint detection, deblending, or photometric and shear measurement.

\section{Conclusion}\label{sec:conclusion}\label{conclusions}

Surveys like the Roman HLIS and the Rubin LSST will revolutionize our understanding of the Universe. They will enable incredible advances in the cosmological study of dark matter and dark energy, in our understanding of astrophysics, and in time-domain studies like supernovae and exoplanets. However, with such statistical power comes exacting requirements on  systematic control. Some of these potential challenges can be mitigated through a joint pixel-level study of the data sets, enabling even more powerful control of systematics and statistical power -- the surveys together are much more powerful than either on its own. Two of these potential benefits -- more precise resolution from overlapping space-based imaging  and a much wider combined wavelength range, spanning from 0.3--2.0 $\mu$m -- can impact a huge range of potential science cases, but in particular, would be game-changing for wide-field cosmology.

The unprecedented volume and complexity of these imaging data sets, however, make combining the information between them in an optimal and unbiased way an equally daunting challenge. Doing so will require significant effort over many years to develop and test algorithms and methods. To validate these approaches, high-fidelity and realistic synthetic survey data from both observatories will be necessary. In this work, we take the first step to producing this kind of resource for the community, combining efforts in the LSST DESC and Roman Cosmology SIT communities to create such a joint synthetic imaging data set.

We have presented and validated 20 deg$^2$ of overlapping synthetic imaging for both the Roman Space Telescope and Rubin Observatory.
This data represents five years of simulated observations of the mock LSST DESC DC2 universe. We have simulated for the first time the detailed physics of the Roman Sensor Chip Arrays derived from lab measurements using the flight detectors for the Roman mission. The simulated imaging and resulting pixel-level measurements of photometric properties of objects are validated in several ways. We have also validated the resulting data from these two image simulation pipelines against one another in overlapping bandpasses, providing a significant cross-check for both pipelines.

As a demonstration of the utility of these matched synthetic surveys, we have used these simulations to explore the relative fraction of unrecognized blends expected in LSST images, finding that 20-30\% of individual objects brighter than $i\sim 25$ in simulated LSST images at five-year depth can be identified as multiple objects in full-depth simulated Roman HLIS images. We have also explored the statistical power at the summary statistic-level of even 20 deg$^2$ of these deep synthetic surveys. Other initial studies at the pixel-level we expect to pursue using these simulations in future work include tests of photo-$z$ improvement through forced photometry in Rubin LSST at Roman detections and using these matched simulations as a test-suite for current multi-resolution deblending approaches.

This work establishes the power of these joint synthetic imaging datasets to explore joint pixel-level processing and calibration to improve deblending of ground-based data like LSST, using wide-field space-based observations like from Roman or Euclid. These simulations provide a unique testing bed for the development and validation of such joint pixel-level analysis techniques for imaging data sets in the second half of the 2020s. Finally, this effort lays the groundwork for future developments to enable easier cross-survey simulation.

\section*{Acknowledgements}

This work was supported by NASA award 15-WFIRST15-0008 as part of the Roman Cosmology with the High-Latitude Survey Science Investigation Team\footnote{\url{https://www.roman-hls-cosmology.space}} and by NASA under JPL Contract Task 70-711320, ``Maximizing Science Exploitation of Simulated Cosmological Survey Data Across Surveys.'' RM acknowledges partial support for this effort from the Department of Energy grant DE-SC0010118. HA was supported by Leinweber Postdoctoral Research Fellowship and DOE grant DE-SC009193. MI acknowledges that this material is based upon work supported in part by the Department of Energy, Office of Science, under Award Number DE-SC0022184 and a SPIRe award at UT-Dallas.

This paper has undergone internal review by the LSST Dark Energy Science Collaboration. The internal reviewers were Matt Becker, Josh Meyers, and Peter Melchior.

The DESC acknowledges ongoing support from the Institut National de
Physique Nucl\'eaire et de Physique des Particules in France; the
Science \& Technology Facilities Council in the United Kingdom; and the
Department of Energy, the National Science Foundation, and the LSST
Corporation in the United States.  DESC uses resources of the IN2P3
Computing Center (CC-IN2P3--Lyon/Villeurbanne - France) funded by the
Centre National de la Recherche Scientifique; the National Energy
Research Scientific Computing Center, a DOE Office of Science User
Facility supported by the Office of Science of the U.S.\ Department of
Energy under Contract No.\ DE-AC02-05CH11231; STFC DiRAC HPC Facilities,
funded by UK BIS National E-infrastructure capital grants; and the UK
particle physics grid, supported by the GridPP Collaboration.  This
work was performed in part under DOE Contract DE-AC02-76SF00515.

We thank Jeff Kruk, Greg Mosby, Bernard Rauscher, and Ed Wollack for useful comments and suggestions.

This work used results from the {\sc Solid-waffle} detector characterization tools; we thank Jenna Freudenburg and Anna Porredon for their contributions to that package.

This paper used resources at the Duke Computing Cluster and the Ohio Supercomputer Center. Plots in this manuscript were produced partly with \textsc{Matplotlib} \citep{Hunter:2007}, and it has been prepared using NASA's Astrophysics Data System Bibliographic Services.

The SCA data files are based on data acquired in the Detector Characterization Laboratory (DCL) at the NASA Goddard Space Flight Center. We thank the personnel at the DCL for making the data available for this project.

Part of the research described in this paper was carried out at the Jet Propulsion Laboratory, California Institute of Technology, under a contract with the National Aeronautics and Space Administration.

The contributions from the primary authors are as follows.
MT was lead writer and developer of Roman image simulation framework.
CL contributed to the roman image simulation and the following scientific discussions. In particular, he simulated the Roman detector non-idealities based on the detector characterization and modeling and wrote the corresponding section for it.
AP contributed to the blending analysis and interpretation of results; wrote the section on the blending analysis.
CH led the processing of the Roman SCA files and the Roman observing sequence, and worked closely with several of the other authors to ensure these were being used correctly in the simulations. He made major contributions to writing those sections of the paper.
RM contributed to simulation design, analysis, and interpretation of results, and wrote paper sections about DC2 simulations and processing.
MJ wrote a lot of the GalSim code that was used in the simulation, including a fair amount specifically to enable this sim. Also helped design/troubleshoot the simulation driver code by MT. Small amount paper writing/editing.
AC contributed to Roman image simulation and detector characterization pipelines and sub-section on statistical power.
JG led investigation exploring charge diffusion and quantum yield in Roman detectors, including code to generate the detector physics models used in the Roman image simulation.
MH contributed to developing and testing detection and photometric measurement algorithms on the Roman simulated images.
BS contributed the difference imaging analysis (e.g. figure 8). Selected the field, ran the subtraction code, combined the panels.
MY contributed to the discussions of the updates to the coadd method, Roman image simulation development, and general paper comments and discussions.
HA contributed the description of the LSST survey and survey strategy (in intro; section 3.2); provided detailed feedback on an earlier draft of the paper.
JC contributed to the design and implementation of the DC2 simulations, including imSim development, instance catalog generation, image processing, and output catalog validation. He consulted with Troxel on use of the DC2 instance and truth catalogs for the Roman simulations.
OD led Roman Cosmology with HLS SIT, and contributed to original idea for joint simulation approach and edits of paper.
CW contributed to DC2 design and production particularly in the development of the imSim package. He also contributed text to the paper describing how imSim was used in the DC2 simulations.
TZ contributed to bug-fixing SCA file handling of the Roman PSF module in Galsim, PSF coadds, and provided feedback to Sec.~5.4.

\section*{Data Availability}

Data products will be available for access/download from IPAC\footnote{\url{https://www.ipac.caltech.edu}} upon publication of this paper.

\bibliographystyle{mnras}
\bibliography{short}

\appendix

\section{The Roman High-Latitude Imaging Survey}\label{hlisdef}

This appendix describes the Reference design of the Roman HLIS in greater detail.

At a very high level, the trade space for the HLIS has a few basic degrees of freedom: the choice of filters; the tiling/dithering pattern; and the survey area per unit time (which can be traded against depth). The filter set for the HLIS was designed to allow multiple ellipticity auto- and cross-correlation functions for internal consistency checks, as well as near infrared photometry that can be combined with LSST optical photometry for photo-$z$'s. The blue end of the HLIS filter set was set to 0.92 $\mu$m to avoid a gap with Rubin Observatory's $z$ band (note that Rubin does have a $y$-band, but a preliminary assessment by the Science Definition Team \citep{2013arXiv1305.5422S} showed that $\sim 40$\% of the Roman HLIS source galaxies would not be detected even at 10-year LSST depth in $y$). The need to provide multiple cross-correlations drove Roman to include a minimum of 3 shape measurement filters in the HLIS; we chose the option of using the Roman Y106 band for photo-$z$'s and doing shapes in three filters (J129/H158/F184) spanning the range up to 2 $\mu$m. The Roman plate scale was chosen to be 0.11 arcsec, so that we are in only the weak undersampling regime (pixel size $\le \lambda/D$; this ensures that at least some Fourier modes in the image are not aliased, see, e.g., \citealt{1999PASP..111..227L}) at J129 band (central wavelength 1.29 $\mu$m).

\begin{figure*}
\includegraphics[width=\textwidth]{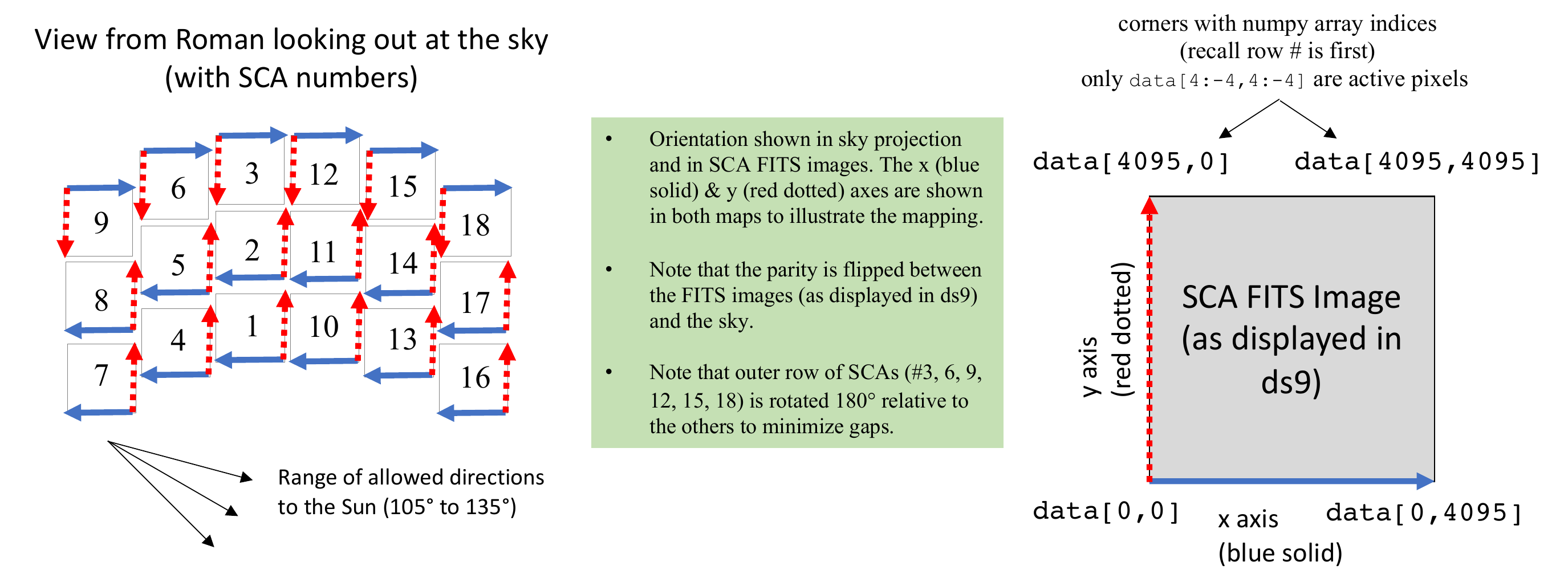}
\caption{\label{fig:SCACoords}The mapping of the Roman focal plane to the FITS files. The ``pawprint'' shape of the 18 SCAs and their numbering is seen in the focal plane layout at left. The 2D SCA image is shown at right in ds9 format (with Python indices overlaid). 
}
\end{figure*}

\begin{figure*}
\includegraphics[width=6.5in]{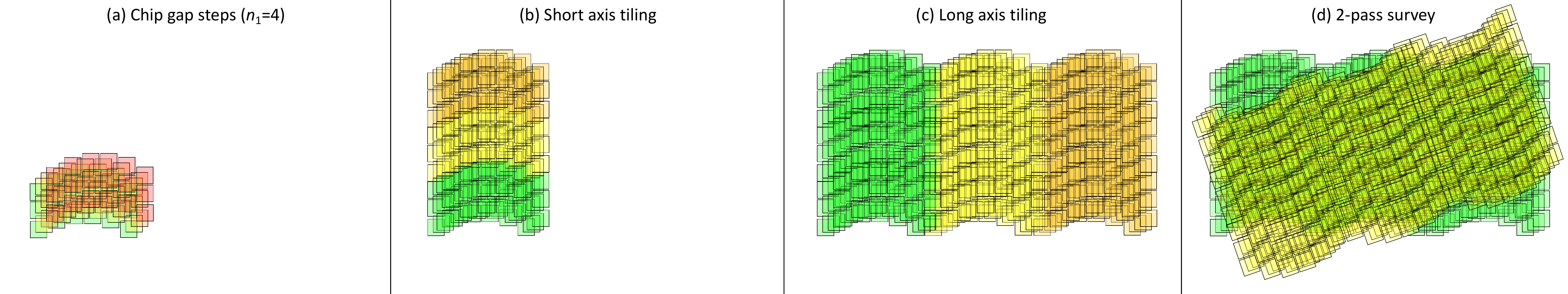}
\caption{\label{fig:imaging_tiling}The sequence of steps in the tiling strategy, from the inner {\tt for} loop (a) to the outer {\tt for} loop (d) described in Sec.~\ref{surveys-roman}. Rotational dithering enters when beginning a new pass in the outer {\tt for} loop (d) and during each filter change.}
\end{figure*}

The basic survey observing pattern was constrained by the field of view layout. An on-axis three-mirror telescope with one or more folds and an accessible exit pupil for stray light control can provide an annular corrected and unobstructed field of view \citep{1977ApOpt..16.2074K}. The 18 SCAs  \citep{2020JATIS...6d6001M} that make up the Roman Wide Field Instrument focal plane are arranged in a $6\times 3$ ``pawprint'' pattern that fits in this annulus (Fig.~\ref{fig:SCACoords}), with chip gaps based on the packaging of the SCAs. The arced pattern can be tessellated to cover a wide survey area. This suggests a scheme for tiling the sky with four nested {\tt for} loops, as shown in Fig.~\ref{fig:imaging_tiling}:
\newcounter{hlisfor}
\begin{list}{(\alph{hlisfor})\ }{\usecounter{hlisfor}}
\item The ``inner'' {\tt for} loop is a set of $n_1$ translational dither steps, sufficient to cover the chip gaps. These are in a diagonal direction so as to not overlay either horizontal or vertical gaps. This step was also designed to dither over large cosmetic defects found in some SCAs. We now know the layout of the most significant large defects from lab tests (although the individual pixel mask may vary during flight, e.g., as seen on HST WFC3-IR; \citealt{2019wfc..rept....3S}); the simulation tools described in this paper will help us optimize the exact choice of dither step.
\item The next loop tiles the pawprint on the short-axis direction, making a strip. We found on the curved sky that placing this before the long-axis step led to higher efficiency.
\item The next loop tiles the pawprint on the long-axis direction into contiguous 2D coverage of a large area to make a ``pass.''
\item The ``outer'' {\tt for} loop consists of $N_{\rm pass}$ passes over the survey footprint, each at a different roll angle. Multiple passes provide an opportunity for internal relative calibration using the survey data itself, which has proven valuable in past wide-field optical surveys \citep[e.g.][]{2008ApJ...674.1217P, 2012ApJ...756..158S}. We have chosen $N_{\rm pass} = 2$ for the wide survey, which is the minimum to provide this internal redundancy: since the Roman HLIS already has significant overheads (21\% in the current Design Reference Mission\footnote{Available on the Roman Project Website; document number RST-SYS-DESC-0073}.), we have not increased it further.
\end{list}
In the present simulated observing plan, the choice of $n_1$ varies from filter to filter depending on the sampling needs. In the J129 filter, we will be doing shape measurement with the smallest PSF and thus we have $n_1 = 4$ so that most of the area gets 6 observations (after accounting for gaps). In the other filters, we have $n_1=4$ for the first passes and $n_1=3$ for the second.

The orientation and scheduling of the passes are constrained by the Roman Sun angle constraint. The solar array is pointed perpendicular to the telescope boresight, and the elongation angle $\varepsilon$ from the boresight to the Sun is constrained to lie between $54^\circ$ and $126^\circ$. Larger elongation angles are preferred to reduce zodiacal background, especially at lower Ecliptic latitude. In Roman's orbit around the Sun-Earth L2 point, the Earth and Moon are on the same side of the spacecraft as the Sun, leading to no additional Earth observing constraint and only occasional Moon observing constraints (the Moon can appear up to $47^\circ$ from the Sun). Regions of sky within 36$^\circ$ of the Ecliptic poles can be observed all year round (the two ``Continuous Viewing Zones'' or CVZs), whereas regions closer to the Ecliptic can be observed during two ``seasons'' each year of length
\begin{equation}
t_{\rm season} \approx \frac{6\,\rm months}{90^\circ}\times \sin^{-1} \left(\frac{\cos 54^\circ}{\cos\beta}\right),
\end{equation}
where $\beta$ is the Ecliptic latitude and the equation is approximate since there are small corrections associated with the eccentricity of Earth's orbit. Since the solar array is fixed to the spacecraft and was placed at a 120$^\circ$ angle to the orientation of the camera (see Fig.~\ref{fig:SCACoords}) to accommodate the mechanical constraints of the inherited 2.4~m telescope, requirements from power, stray light, and thermal control needs lead to a requirement that the observatory is rolled no more than 15$^\circ$ from the direction to the Sun. This leads to a constraint on the possible roll angles of the telescope when more than $\sim 1^\circ$ outside of the CVZ (Eq.~1 of \citealt{2022ApJ...928....1W}). Furthermore, whether inside or outside the CVZ, there is a joint constraint on the elongation and the Ecliptic position angle\footnote{This is defined so that the ``top'' of the pawprint -- the WFI Local +Y in Fig.~\ref{fig:SCACoords} -- is at an angle $\psi$ from the direction to the North Ecliptic Pole, with the sign convention that Ecliptic East is $+90^\circ$.} $\psi$ of the telescope:
\begin{equation}
\psi = \left\{ \begin{array}{rc} 210^\circ+ \sin^{-1} \left( \tan\beta \cot\varepsilon \right)
 & {\rm leading} \\ 30^\circ- \sin^{-1} \left( \tan\beta \cot\varepsilon \right) & {\rm trailing} \end{array}\right\}  \pm 15^\circ,
\label{eq:psi}
\end{equation}
where the $\pm 15^\circ$ is the allowed range of rolls relative to the Sun, and ``leading'' and ``trailing'' refer to cases where the telescope is pointed forward or backward in the Earth's orbit, respectively.\footnote{If $\varepsilon<|\beta|$ or $\varepsilon>180^\circ-|\beta|$, the argument of the arcsine in Eq.~(\ref{eq:psi}) is outside the valid range of $-1$ to $+1$; in this case, that field can never be observed at that elongation angle from the Sun.} The orientation of the tiling pattern is pre-selected so that Y106/J129/H158 bands are observed at lower zodiacal levels (F184 has significant thermal background), and so that in each filter one of the passes is leading and the other is trailing. The orientation of the tiling is manually adjusted until the simulated scheduling run is successful. In general, we have found that in order to get a consistent scheduling solution, we have to break the survey footprint into a few (here 8) ``sectors,'' each with a somewhat different orientation. Different orientations are also chosen for each filter so that we do not repeat the exact same set of pointings.

The Reference HLIS includes a set of 3 deep fields, each circular and $3.5^\circ$ in diameter to match the Rubin field of view, that have lower noise and better sampling than the remainder of the survey. These are covered with $N_{\rm pass} = 10$ passes. Right now these should be thought of as placeholders for the actual deep field plan. One of these is placed in the Southern CVZ and the passes are distributed in roll angle; the others are outside the CVZ and the roll angles are distributed in a ``bowtie'' fashion since the Sun angle constraints do not allow every possible roll. These deep fields are not simulated in this paper.

The choice of exposure time reflects a depth vs.\ area trade. For very long exposures, the limiting flux scales as $F\propto t_{\rm exp}^{-1/2}$; since the galaxy number counts are shallow ($d\ln N/d\ln F \approx -0.8$ as measured in CANDELS 1.6 $\mu$m data; \citealt{2011ApJS..197...35G, 2011ApJS..197...36K}), a faster survey observes more galaxies overall, in addition to reducing cosmic variance. However, very short exposures are limited by slew and settle overheads and read noise. The Roman Science Requirements Document\footnote{Available on the Roman Project Website; document number RST-SYS-REQ-0020C\_SRD.} was written around a choice of 2000 deg$^2$ for the Reference Survey, based on a preliminary optimization by the Science Definition Team \citep{2015arXiv150303757S}. This leads to an exposure time of 139.8 s. We have deferred re-optimization of this area until after the full focal plane test for Roman, when we expect to have a much better understanding of the ``as-built'' read noise properties.

The final degree of freedom for the tiling is the choice of survey footprint. The Roman Reference HLIS is designed to be contained in the LSST footprint; to stay away from the Ecliptic to reduce zodiacal background (we place sector boundaries at least 19$^\circ$ away); and to avoid low Galactic latitudes or high dust column. In addition, it is desirable that at least some of the footprint be accessible from Northern Hemisphere facilities, both for follow-up of Roman HLIS objects and for overlapping survey programs with, e.g., Subaru. The only way this can be done with a contiguous survey area is to choose an RA range where the Ecliptic is relatively far north (2--10$^{\rm h}$) and that avoids the Galactic Plane crossing the Equator (7$^{\rm h}$). These constraints naturally lead to placement centered at $\sim 3^{\rm h}$ right ascension. The Reference survey footprint is shown in Fig.~2 of \citet{2022ApJ...928....1W}.

We reiterate that while the Roman hardware parameters have been chosen, \textit{all of the survey strategy choices in the Reference Survey are subject to further revision, including major changes of choice of footprint, depth vs.\ area, and exploration of multi-tiered options}. We expect the simulation tools described here to inform those further discussions.

\def\arraystretch{1.4}
\setlength{\tabcolsep}{4pt}
\begin{table*}
\caption{Summary of the v1 SCA files describing model parameters of various physical effects within the detectors. Dimensions are shown in the numpy.shape convention. In most cases, if the binning is $n$ pixels, the last two axes have dimensions $4096/n$ and correspond to the $y$ and $x$ indices of the binned pixels. The first axis (if 3D) or first two axes (if 4D) can be used for other information, such as wavelength dependence, order of polynomial coefficient, or component of a kernel (details in text of Appendix~\ref{detectors}). The exceptions are the starred (*) dimensions which contain 2D arrays for easier display in ds9 but have the same content as another HDU (e.g., IPCFLAT is a display-friendly organization of IPC).}
\label{table:sca}
\begin{center}
\begin{tabular}{clcccl}
\hline
\hline
HDU & Contents & Type & Binning & Dimensions & Status \& source data \\
\hline
PRIMARY & Empty & PrimaryHDU & N/A & N/A & Present (FITS standard) \\
SOURCES & Record of input files & BinTableHDU & N/A & 1 column & Present \\
RELQE1      &       Per-pixel relative QE & ImageHDU, float32 & 1 pix & (4096,4096) & Acceptance flats \\
RELQE2      &       Wavelength-dependent relative QE & ImageHDU, float32 & 64 pix & (50,64,64) & Placeholder \\
 QYIELD      &       Quantum yield & ImageHDU, float32 & 128 pix & (50,32,32) & Placeholder \\
 CHRGDIFF      &       Charge diffusion & ImageHDU, float32 & 128 pix & (3,32,32) & Placeholder \\
  BFE      &       Brighter-fatter effect & ImageHDU, float32 & 128 pix & (5,5,32,32) & Acceptance, solid-waffle \\
 BFEFLAT      &     & ImageHDU, float32 & 128 pix & (160,160)* & Same as BFE, for 2D display in ds9 \\
  PERSIST      &       Persistence & ImageHDU, float32 & 1 pix & (6,4096,4096) & Acceptance persistence data \\
 DARK      &       Dark current & ImageHDU, float32 & 1 pix & (4096,4096) & Acceptance, combined short+long darks \\
CRNL     &       Charge-rate nonlinearity & ImageHDU, float32 & 1 pix & (1,4096,4096) & Placeholder \\
  SATURATE     &       Saturation & ImageHDU, int32 & 1 pix & (4096,4096) & Acceptance, full well data \\
  CNL     &       Classical Nonlinearity & ImageHDU, float64 & 1 pix & (3,4096,4096) & Acceptance, full well data \\
  BURNIN     &       Burn-in & & & & Not yet implemented \\
  IPC     &       Interpixel capacitance & ImageHDU, float32 & 8 pix & (3,3,512,512) & Acceptance, single pixel reset \\
  IPCFLAT     &   &  ImageHDU, float32 & 8 pix & (1536,1536)* & Same as IPC, for 2D display in ds9 \\
  VTPE     &       Vertical trailing pixel effect & & & & Not valid in v1 files \\
  BADPIX     &       Pixel bitmask & ImageHDU, uint32 & 1 pix & (4096,4096) & Flagged while constructing other HDUs \\
  READ     &       Read noise & ImageHDU, float64 & 1 pix & (3,4096,4096) & Acceptance darks \\
  GAIN     &       Gain & ImageHDU, float32 & 128 pix & (32,32) & Acceptance, solid-waffle \\
  BIAS     &       Bias & ImageHDU, uint16 & 1 pix & (4096,4096) & Acceptance darks \\
\hline
\hline
\end{tabular}
\end{center}
\end{table*}

\section{Simulating the Roman detectors}\label{detectors}

The physics associated with converting photons into a digitized signal within the Roman detector system are described in a set of 18 FITS files (one for each SCA), with an HDU describing each non-ideal effect (including, where appropriate, its spatial variation). These are inputs to the image simulation pipeline. The file structure is summarized in Table \ref{table:sca}, while individual physical effects that we model are described below. This appendix describes the first version (v1) of these files, which is based on a combination of data from Roman acceptance testing, and some placeholders. We anticipate regular updates to these files to improve their fidelity as more calibration data becomes available, both based on later ground testing at a higher level of integration and ultimately including in-flight calibration data as well. These updates will likely include improvements to the modeling of existing effects, as well as new effects if they are discovered.

The v1 files were intended primarily for development and testing of simulation infrastructure (including calibration tools). Caution should be used before inferring specific aspects of Roman performance from these files, especially since the SCA acceptance testing differs from the flight configuration in some important respects, and some effects are not included. The most important caveats and known issues are as follows:
\begin{list}{$\bullet$}{}
\item Some HDUs are placeholders (these are identified in the last column of Table \ref{table:sca}).
\item Some physical effects in the SCAs are known to exist but are either not yet implemented (e.g., burn-in; see \S2.7.2 of \citealt{2020JATIS...6d6001M}) or there was not a valid description in the v1 files (e.g., vertical trailing pixel effect; see \citealt{2020PASP..132g4504F}).
\item Most of the HDUs in the v1 files are based on acceptance test data for the SCAs, with a laboratory (Generation 3 Leach) controller. In the flight configuration, the ACADIA controller \citep{2018SPIE10709E..0TL} will be used. Some properties that are intrinsic to the SCA (e.g., cosmetic defects, IPC) are expected to be similar. Properties that depend on how the signal is digitized (e.g., gain, classical non-linearity) as well as noise properties will be different in the flight configuration. There is now test data of each SCA with an ACADIA controller, but it was not available in time for the processing for the v1 files; we plan to incorporate this data in the v2 and later detector files.
\item Further adjustments to the operational parameters of the detector system, difference in stability levels in the flight environment, or degradation due to aging or radiation damage may lead to changes even after launch. We anticipate switching from static detector effect files to a time series of such files (cadence to be determined) once there is a sufficient baseline of on-orbit data to warrant this.
\item The noise correlations in flight will differ with the ACADIA controller and depend on how reference pixel correction will be performed. In particular, there is an effort underway to apply optimal linear reference pixel correction \citep{2022JATIS...8b8002R} to data acquired with the ACADIA and incorporate the H4RG-10's reference output as an additional template. Incorporation of a correlated noise model and assessment of its impact are deferred to a future version of this simulation package and the v2 SCA files.
\item Because the darks and flats have not been IPC-corrected, IPC artificially acts twice on dark and flat features in the final simulated images using the v1 files. This phenomenon is especially important for the ``footprint'' of defects such as hot pixels.
\item While the detector files describe ``permanent'' cosmetic defects such as dead pixels, transient events such as cosmic ray strikes are not included in the static detector files. At a typical interplanetary rate of $5.5$ cm$^{-2}$ s$^{-1}$, we expect $\sim 1.3\times 10^4$ cosmic ray events per SCA per 140 s HLIS exposure (see \citealt{2016PASP..128c5005K} for a Roman-specific discussion, and note that at the Sun-Earth L2 Lagrange point we do not have the trapped electron flux).
\end{list}

We now turn to the modeling and construction of HDUs for each of the detector effects. The inputs make extensive use of the solid-waffle toolkit \citep{2020PASP..132a4501H, 2020PASP..132a4502C, 2020PASP..132g4504F, 2022PASP..134a4001G}.

\begin{figure}
\begin{center}
 \includegraphics[width=\columnwidth]{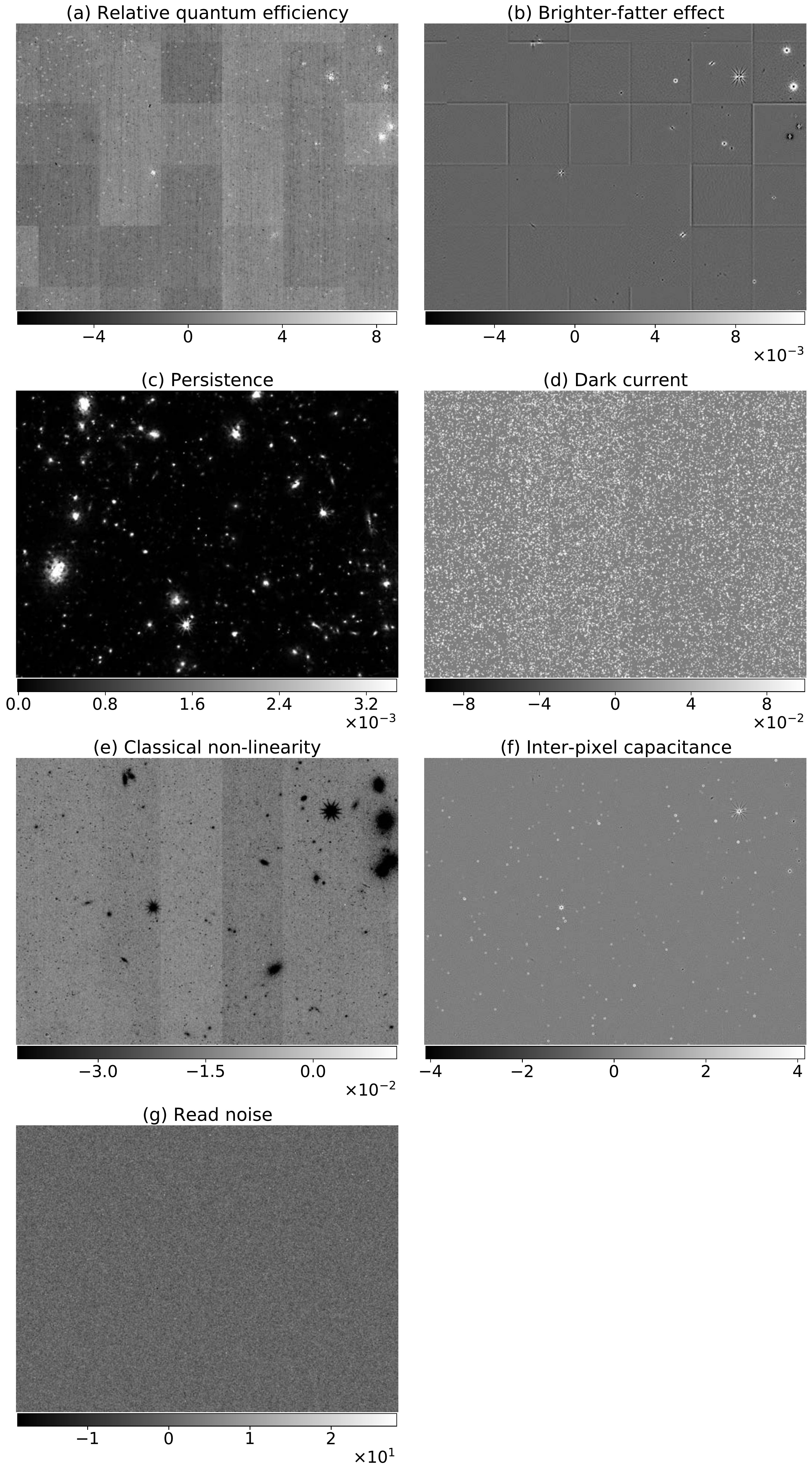}
\end{center}
 \caption[]{Difference images (physical effect on minus physical effect off) showing the effects of various detector physics listed in Appendix \ref{detectors}. The images are plotted using dynamic zscale interval on 800x600 pixels with native Roman pixel scale.
 \label{fig:detector_effect}}
 \end{figure}

\begin{itemize}
    \item \textbf{Relative quantum efficiency}:  The relative quantum efficiency (relative QE) is the spatial variation of throughput that includes contributions from anti-reflection coating, photon conversion into electron-hole pairs, and charge collection in the SCAs.  Relative quantum efficiency of 1 corresponds to the assumed throughput without being affected by the effects considered. Multiplying the pixel-to-pixel map of relative QE yields the output of the effect. The relative QE is modeled with two separate HDUs, RELQE1 and RELQE2, with RELQE1 having full spatial resolution (individual pixels) but no wavelength dependence, and RELQE2 having wavelength dependence but binned pixels (appropriate for large-scale variations in the anti-reflection coating). Their effects multiply.

For the v1 files, we use acceptance flats to construct the RELQE1 map, and a placeholder for the wavelength dependence (RELQE2). The flats have not been corrected for the illumination pattern in the acceptance test setup, so they contain large-scale gradients that are not representative of what is expected in flight. The final calibration of the large-scale flat field response will be performed with astronomical sources. Panel (a) of Fig. \ref{fig:detector_effect} shows the effect of multiplying the map of relative quantum efficiency on the image.

    \item \textbf{Brighter-fatter effect}:   The brighter fatter effect is a non-linear effect that deflects charges due to the electric field built by the accumulated charges, thus making the profiles of more luminous objects appear broader. This effect exists in both CCD and CMOS detectors and is typically at the percent level for bright but unsaturated objects. This effect can also be understood effectively as change in pixel area and pixel boundaries. The brighter-fatter effect is defined in terms of the Antilogus coefficient kernel \citep{2014JInst...9C3048A} of total pixel area change in the detector effect characterization file. The effective pixel area $\mathcal{A}_{i,j}$ for pixel at position $[i,j]$ is described as
    \begin{equation}
        \mathcal{A}_{i,j} = \mathcal{A}_{i,j}^0 \left[ 1 + \sum_{\Delta i, \Delta j}a_{\Delta i, \Delta j} Q(i+\Delta i, j+\Delta j) \right],
\label{eq:A-bfe}
    \end{equation}
    where $\mathcal{A}_{i,j}^0$, $Q$ and $a_{\Delta i, \Delta j}$ are the original pixel area, charge of neighboring pixels and the Antilogus coefficient kernel \citep{2020PASP..132a4501H}. The Antilogus coefficient kernel $a_{\Delta i, \Delta j}$ is provided in the SCA characterization file with $\Delta i$ and $\Delta j$ in the range of $[-2, 2]$. The kernel for the total pixel area, however, is not sufficient. Image simulation of the brighter fatter effect requires the shift of the four pixel boundaries. In the implementation, we solve for the four boundary shift components (top, bottom, left, right) from the kernel of total pixel area by assuming several symmetric constraints \citep{2015A&A...575A..41G,2015JInst..10C5032G}. The four boundary shift components of the area kernel is then used to simulate brighter-fatter effect.

The BFE kernel was estimated by the flat field correlation method on acceptance flats using the solid-waffle code \citep{2020PASP..132g4504F}, which uses multiple non-destructive reads to separate the contributions from IPC and BFE.
Results are presented in spatial superpixels, so the BFE HDU has dimensionality of (5,5,32,32) where the first two axes are the kernel  indices ($\Delta j$ and $\Delta i$ in Eq.~\ref{eq:A-bfe}) and the second are spatial superpixel indices.
    Panel (b) of Fig. \ref{fig:detector_effect} shows the difference image of brighter-fatter effect. At the center of bright sources, we are able to see charge decrease at center and charge outflow around the center. Since the Antilogus coefficient kernel is binned in 128 pixels, there is slight discontinuity at the boundary.

    \item \textbf{Persistence}:    Persistence is a memory effect in which a bright source observed in one image results in a fading ``after-image'' in the same pixels in subsequent exposures. It results from charge trapping and release at defects in the Hg$_{1-x}$Cd$_x$Te semiconductor \citep[e.g.][]{2008SPIE.7021E..0JS}, although multiple trap populations are involved and a complete model remains elusive.
     In the simulation, the persistence is treated as an effective\footnote{In the H4RG-10, the current or charge through the photodiode is not actually measured. Rather the voltage on the p-type side of the diode is transferred through a source follower and ultimately to the ADC. The voltage change $\delta V$ produced per elementary charge released depends on the location of the trap (see, e.g., \citealt{2020JATIS...6d6001M}, \S2.7.1, and references therein), but when we convert the slope of the ADC output (in DN/s) into a current image (in e/s) we use the conversion appropriate for a hole that crosses from the n to p-type side. Thus the addition to the image due to release of trapped charges is not necerssarily the true rate of release of charge -- hence the ``effective'' current.} current of the form
    \begin{equation}
        I^{\rm pers}_{i,j} =  f_{i,j}(Q_{i,j}) / t,
    \end{equation}
    where $I^{\rm pers}_{i,j}$, $f_{i,j}$ and t are the persistence current at pixel $[i,j]$ in unit of ($e^-$/pixel/second), the charge-dependent persistence response function at $[i,j]$ and the time since the end of the specific exposure (assuming linear decay over time). Previous generations of detectors have also shown a dependence on exposure time \citep[e.g.][]{2015wfc..rept...15L}, however we have not yet incorporated this into the model.

The function $f$ in the v1 files is constructed from a set of measurements at 6 different illumination levels (with timing shown in Fig.~13 of \citealt{2020JATIS...6d6001M}). The form of the function can vary significantly even within an SCA, with some patches showing a step-like function around the saturation level. In the simulation, persistence $f_{i,j}(Q_{i,j})$ as a function of charge is implemented as linear interpolation between 6 charge levels where lab measurement of persistence was made. Similar to \cite{2022MNRAS.512.3312L}, we choose a linear decay of the persistence function $f_{i,j}$ for low level charges, as this fits the low (below half full well) data much better than an extrapolated analytic function. An example of persistence image from an earlier exposure is shown in (c) of Fig. \ref{fig:detector_effect}.

    \item \textbf{Dark current}:    Dark current is the leakage current of the system even when there is no input signal. This current is generally generated from thermal effects in the Hg$_{1-x}$Cd$_x$Te photodiodes. In flight, there will be an additional contribution from internal thermal background in the wide field instrument. It is specified as the average number of electrons per unit time and is implemented as a Poisson noise. The v1 dark current maps were synthesized from long (2 hr) darks obtained during acceptance testing. For hot pixels that saturate during these darks, we use the 150 s long noise data; and for very hot pixels, we use the first two reads to determine the dark current.

We expect the hot pixel population to evolve in flight as individual defects are created by radiation damage \citep[e.g.][]{2004SPIE.5167..258P, 2019wfc..rept....3S}. Dark current maps will be regularly updated based on dark frames taken during the mission; the current baseline is to do this weekly.\footnote{See the Design Reference Mission, available on the Project website at: https://asd.gsfc.nasa.gov/romancaa/docs2/RST-SYS-DESC-0073-\_D10.xlsx}

    \item \textbf{Saturation}:   The saturation level is the maximum number of elementary charges per pixel for which our nonlinearity model is valid. This is measured from full well data in DN and converted to elementary charges using the solid-waffle gain measurement. The  saturation level is typically $\sim 1.1\times 10^5$ e and varies by around $4\%$ across pixels. In the simulation, we perform saturation by clipping the charge of each pixel at the saturation level.

    \item \textbf{Classical non-linearity}:    The response of the H4RG-10 detector arrays is a non-linear function of the collected charge. This effect occurs in a p-n junction since the capacitance depends on the amount of charge stored (and hence the size of the depletion region); additional non-linearities can then arise in the sequence of source followers and the ADC.
Note that this total-signal-dependent non-linearity (or classical non-linearity, CNL) is distinct from the count-rate non-linearity (CRNL) where the photon-generated charge of fixed photons is dependent on the photon rate.
The current model uses a 4th order polynomial:
    \begin{equation}
        Q^{\rm CNL} = Q - \beta_2 Q^2 - \beta_3 Q^3 - \beta_4 Q^4.
    \end{equation}
Fitting a polynomial to the graph of the signal (in DN) versus time (in frame number) in a flat field with constant illumination enables one to measure the combinations $\beta_m g^{m-1}$, where $g$ is the gain. The solid-waffle gain measurement (see below) is used to convert these into $\beta_m$ coefficients. Because the $\beta_m g^{m-1}$ measurement can be performed on a per-pixel basis, whereas the gain measurement is noisy and has to be binned (we use $128\times 128$ super-pixels), the $\beta_m$ maps have the super-pixel pattern superposed. One can also see the 128-column-wide bands in Fig.~\ref{fig:detector_effect}(e), which correspond to the 32 readout channels.

\begin{figure}
\begin{center}
\includegraphics[width=\columnwidth]{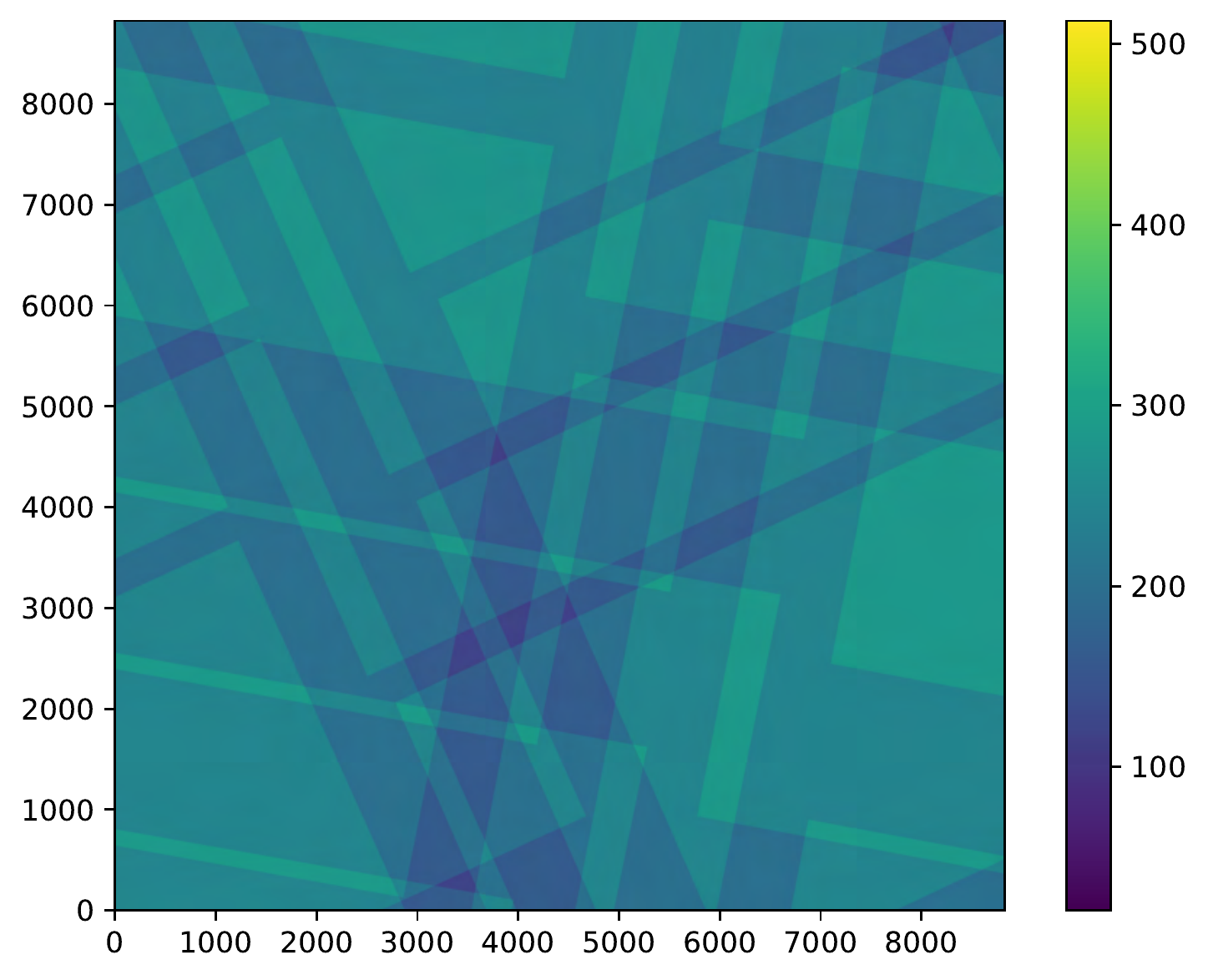}
\includegraphics[width=\columnwidth]{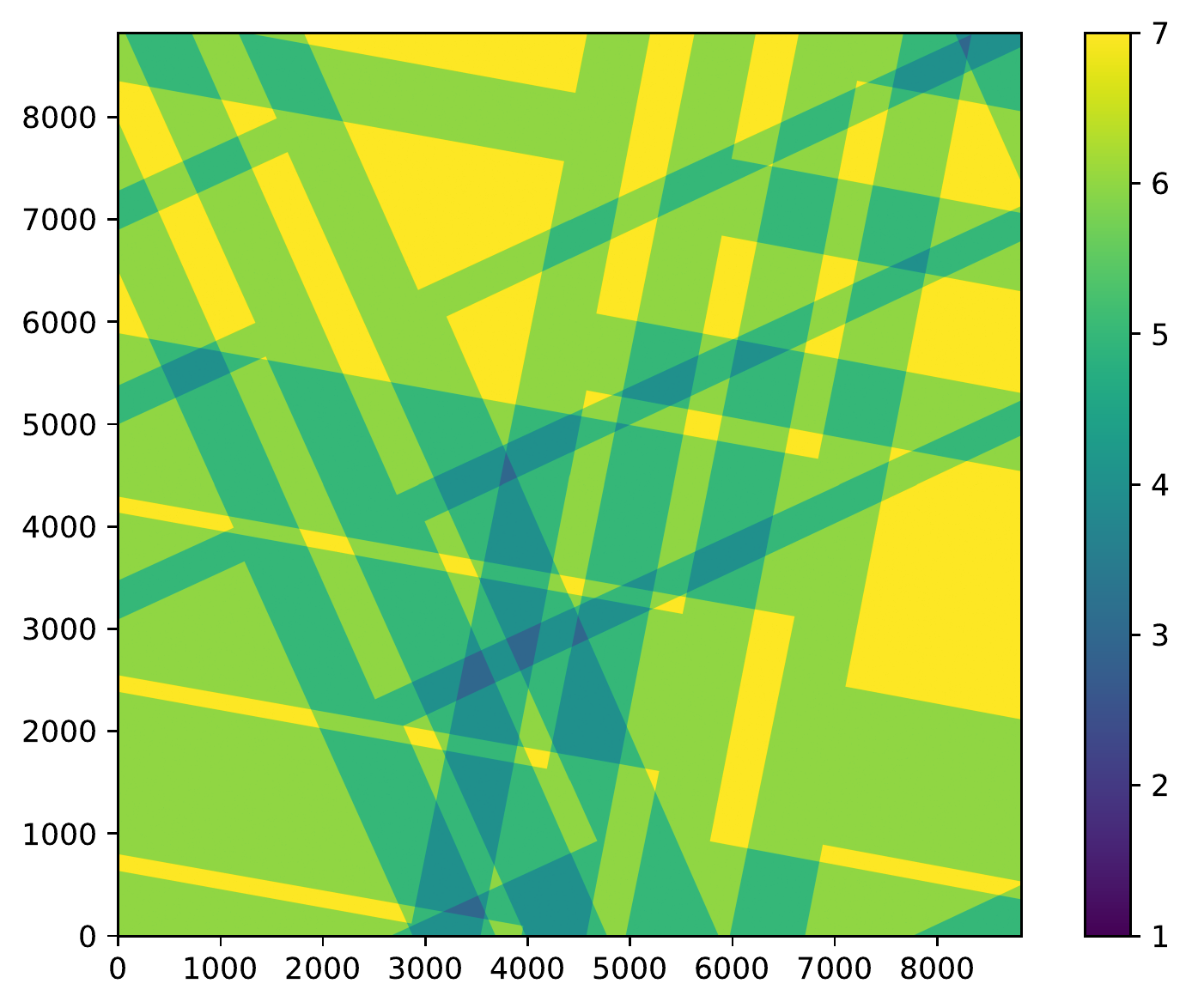}
\end{center}
\caption[]{Example Roman coadd weight (top) and number of images (bottom) maps for exposures using the H158 filter at a random region within the simulated sky area. There is correlated variance in the weight maps due to the drizzling algorithm used on scales of a few pixels (about 0.2 arcsec) -- a noise image capturing these effects and other sources of correlated noise in the detectors are provided with each coadd image.
\label{fig:weight}}
\end{figure}

The classical non-linearity is in the CNL HDU, with shape (3,4096,4096); the first axis maps to the order of the polynomial coefficient.

    \item \textbf{Inter-pixel capacitance}:    Inter-pixel capacitance is a source of crosstalk between pixels that originates from the capacitance coupling between adjacent pixels in the source-follower \citep{2016PASP..128i5001K}. The signal at a given pixel is then coupled with neighboring pixels due to the capacitance coupling. This inter-pixel capacitance effect is modeled as a linear effect that is described as a convolution of a $3\times 3$ kernel with the image, leading to blurring effect on images. The corrected charge after including inter-pixel capacitance follows
\begin{equation}
        Q_{i,j}^{\rm IPC} =  \sum_{\Delta i,\Delta j} K_{i-\Delta i, j-\Delta j}(\Delta i, \Delta j) Q_{i-\Delta i, j-\Delta j},
\end{equation}
where $K(\Delta i, \Delta j)$ is the IPC kernel documented in the SCA file.

The IPC is measured from acceptance test data using the single pixel reset method \citep{2008SPIE.7021E..04S}. This is a purely electronic method in which the reset line is used to set an individual pixel to a specified voltage while its neighbors are disconnected (i.e., have constant charge). The change in signal on the neighbor pixels allows us to infer the IPC. The resets are performed on every 8th pixel (in row and in column), so we use $8\times 8$ pixel binning to make a full IPC map. In some cases, we had to pad the IPC map by 1 pixel due to a channel boundary or reference pixel, or interpolate if the pixel that was reset in a given $8\times 8$ patch was masked.

The IPC HDU is a 4D array of dimension (3,3,512,512): the first two axes correspond to the kernel axes $(\Delta j,\Delta i)$, and the second to spatial position on the SCA. The IPCFLAT HDU is 2D version suitable for display in a tool such as ds9.

    An example of IPC effect is shown in (f) of Fig. \ref{fig:detector_effect}, illustrating the smoothing effect of the 3x3 IPC kernel.

    \item \textbf{Read noise}:  The v1 files contain a map of read noise generated from the slope (d[signal]/d$t$) images in 100 dark frames of 55 reads each, acquired during acceptance testing. (This slope noise is in the 3rd slice, {\tt data[2,:,:]} in astropy.io.fits format; the other slices {\tt data[0,:,:]} and {\tt data[1,:,:]} contain a principal component map and the correlated double sample noise, respectively.) The current simulation includes Gaussian read noise. There is significant correlated noise, both in data with the Leach controller and the ACADIA controller. We are planning to include correlated noise in a future version of GalSim (and put the relevant parameters in the v2 files).

    \item \textbf{Gain}: The inverse gain $g^{-1}$ (in DN/e) is the conversion from collected charge (in elementary charges) to observed signal (in data numbers). It includes several pieces of physics, including the effective node capacitance on the pixel (charge to voltage); the gains $dV_{\rm out}/dV_{\rm in}$ of each step in the amplifier chain; and the $\Delta V$ increment of the levels in the ADC. These are not individually measured, rather we use the photon transfer curve method \citep{1981SPIE..290...28M, 1985SPIE..570....7J}, as implemented in solid-waffle \citep{2020PASP..132g4504F}, which relies on the Poisson distribution (variance = mean when measured in units of elementary charges) to determine $g$. Due to signal-to-noise limitations, solid-waffle was run with 128$\times$128 pixel binning.

The gain in the flight configuration will be different from the SCA acceptance test data. However, inclusion of a gain field is necessary in order to have a complete simulation chain, and enables us to test the interface between solid-waffle and GalSim.

\end{itemize}

    Besides the data product that includes the full detector physics models listed above, we also provide a set of simulation that only keep simplified detector effects. In the simple detector physics model, we remove the detector non-idealities and keep only dark current, saturation and read noise. Unlike the full detector physics model that requires full characterization data of various detector physics of 18 SCAs as inputs, the simple model applies the noises with the same set of parameters for pixels and SCAs. The parameters we use for the simple model are: dark current of $0.015 \; (e^-/s/{\rm pixel})$, saturation level of $10^5 \; e^-$ and read noise of $8.5 \; e^-$. With the simple model, we provide simulated images that are closer to calibrated images that are ready for further analyses while a the same time retain the noise levels.

\section{Post-processing of Roman images}\label{app:coadd}

We describe additional details about processing performed on the individual Roman SCA images in this appendix.

\subsection{Coadd images}

Coadding of individual images and background subtraction are performed using \textsc{AstroDrizzle}\footnote{https://drizzlepac.readthedocs.io} v3.2.0 \citep{2002PASP..114..144F}. We calibrate the pixel fraction used for drizzling to the five year Reference HLIS depth, inspecting for the presence of empty pixels in the final images and otherwise choosing a value such that the median-normalized rms of the coadd weight map remains below 0.2. This results in a final pixel fraction of 0.7 and corresponding final pixel scale of 0.0575 arcsec (about half the original pixel scale). We weight the coadded images by exposure time, which is constant for all HLIS images, and drizzle an equivalent set of noise images to produce a properly correlated representation of the coadd noise. We show an example of a coadd weight map and the number of exposures contributing to each pixel in Fig.~\ref{fig:weight}.

The coadd PSF model is not drizzled, but linearly coadded at high resolution. Since each SCA is simulated with a uniform PSF across it, we select PSF model images with an appropriate WCS for each input image in the coadd. These model images use a pixel scale of 0.01375 arcsec ($1/8$ the native Roman pixel scale). The images are interpolated using a \textsc{lanczos5} interpolant, and a coadd PSF of these summed model images is produced for every unique combination of input images, as interpreted from the context array produced by \textsc{AstroDrizzle}. These high-resolution models are then downsampled to a target pixel scale using the same \textsc{lanczos5} interpolant with a function provided as part of the Roman simulation suite.

\subsection{Object detection and photometry}

A detection image is created by taking the median of the four coadd images in each bandpass observed in the Roman HLIS. Object segmentation, centroid determination, and basic photometric fits are performed with \textsc{Source Extractor} \citep{1996AAS..117..393B,Barbary2016}. The detection threshold is set to 2.5$\sigma$, with a minimum pixel area of 5. The deblending thresholds level was chosen to be 48 and the minimum contrast ratio to be 0.05. Basic photometric measurements for detected objects are performed via forced photometry in each bandpass.

\bsp
\label{lastpage}
\end{document}